\documentclass{aa}

\usepackage{natbib}
\bibpunct{(}{)}{;}{a}{}{,} 
\usepackage{graphics}   
\usepackage{graphicx}   
\usepackage{epstopdf}
\usepackage{longtable}   
\usepackage{url}      
\usepackage{bm}        
\usepackage[varg]{txfonts}
\usepackage{pdflscape}
\usepackage{caption}
\usepackage{supertabular}
\usepackage{hyperref}   

\hypersetup{
    bookmarks=true,         
    unicode=false,          
    colorlinks=true,       
    linkcolor=blue,          
    citecolor=blue,        
    filecolor=blue,      
    urlcolor=blue           
}
\begin{document}

\title{Star-planet interactions} 
\subtitle{II. Is planet engulfment the origin of fast rotating red giants?}

\author{Giovanni Privitera\inst{1,3}, Georges Meynet\inst{1}, Patrick Eggenberger\inst{1}, Aline A. Vidotto\inst{1}, Eva Villaver\inst{2}, 
\and
Michele Bianda\inst{3}
}

 \authorrunning{Privitera et al.}

 \institute{Geneva Observatory, University of Geneva, Maillettes 51, CH-1290 Sauverny, Switzerland
\and  Department of Theoretical Physics, Universidad Autónoma de Madrid, M\'odulo 8, 28049 Madrid, Spain
 \and Istituto Ricerche Solari Locarno, Via Patocchi, 6605 Locarno-Monti, Switzerland
}

\date{Received /
Accepted}
\abstract  {
Fast rotating red giants in the upper part of the red giant branch have surface velocities that cannot be explained by single star evolution.}
{We check whether tides between a star and a planet followed by planet engulfment can indeed accelerate the surface rotation of red giants for a sufficient long time in order to produce these fast rotating red giants.}
{Using rotating stellar models, accounting for the redistribution of the angular momentum inside the star by different transport mechanisms, for the exchanges of angular momentum between the planet orbit and the star before the engulfment and
for the deposition of angular momentum inside the star at the engulfment, 
we study how the surface rotation velocity at the stellar surface evolves. 
We consider different situations with masses of stars in the range between 1.5 and 2.5 M$_{\odot}$, masses of the planets between 1 and 15 M$_{\rm J}$ (Jupiter mass), and initial semi-major axis between 0.5 and 1.5 au. The metallicity Z for our stellar models is 0.02.} 
{We show that the surface velocities reached at the end of the orbital decay due to tidal forces and planet engulfment can be similar to values observed for fast rotating red giants. This surface velocity then decreases when the star evolves along the red giant branch but at a sufficiently slow pace for allowing stars to be detected with such a high velocity. More quantitatively, star-planet interaction can produce a rapid acceleration of the surface of the star, above values equal to 8 km s$^{-1}$, for periods lasting up to more than 30\% the red giant branch phase. 
As found already by previous works, the changes of the surface carbon isotopic ratios produced by the dilution of the planetary material into the convective envelope is quite modest. Much more important might be the increase of the lithium abundance due to this effect. However lithium may be affected by many different, still uncertain, processes. Thus any lithium measurement can hardly be taken as a support or as an argument against any star-planet interaction.} 
{The acceleration of the stellar surface to rotation velocities above limits that depend on the surface gravity does appear at the moment as the clearest  signature of a star-planet interaction.}
\keywords{}

\maketitle

\titlerunning{Star-planet interactions}

\authorrunning{Privitera et al.}

\section{Introduction}
\indent It is well known that when stars evolve into the red giant branch after the main-sequence phase, the very large expansion of the envelope imposes very low surface rotation velocities. Indeed, early survey of projected rotational velocities $v \sin i$ \citep{gray81,gray82} showed that giants cooler than about 5000 K are predominantly slow rotators and characterized by $v \sin i$ of a few km s$^{-1}$ (see Fig.~\ref{fig:obs2011}). However, there exists a few percents of red giants presenting much higher $v \sin i$ \citep{fekel93, massarotti08, carlberg11, Tayar2015}.

To explain the high rotation rates of (apparently) single red giants, two kinds of scenarios have been proposed: a first scenario involves mechanism occurring in the star itself. \citet{simon89}
proposed that, at the time of the first dredge-up, the surface could be accelerated by the transfer through convection of angular momentum from the central fast spinning regions to the surface.
\citet{fekel93} explained the high rotation and the high lithium abundance they observed in their sample of red giants as resulting from such a scenario. The dredge-up episode would bring to the surface not only angular momentum but also freshly synthesized lithium. 
It is interesting to underline here that this dredge-up scenario is expected to cause rapid rotation at a particular phase in giant stars evolution, namely when the first dredge-up occurs.
In the large sample studied by \citet{carlberg11}, a clustering of the rapid rotators at this phase (between T$_{\rm eff}$ equal to $\sim$4500 and 5500 K or log T$_{\rm eff}$ between 3.732 and 3.740) is not seen. 
Moreover, we showed in \citet[][accepted for publication in A\&A]{paperI} that the dredge-up actually produces no significant acceleration of the surface and thus cannot be a realistic reason for the high surface rotations that we are discussing here.

A second scenario proposed to explain high rotation rates in giants involves the swallowing of planet \citep{peterson83,siess99I,livio02,massarotti08,carlberg09}. 
The phenomenon of planet/brown dwarf ingestion was studied theoretically by \citet{sandquist98,sandquist02} for Main-Sequence (MS) stars and by \citet{soker84,siess99I,siess99II} and recently by \citet{Passy2012} and \citet{Staff2016} for giant stars. 
\citet{siess99I,siess99II} studied the accretion of a gaseous planet by a red giant and an asymptotic giant branch star (AGB). They considered cases where the planet is destroyed
in the stellar envelope. They focused their study on possible consequences of this engulfment on the luminosity and surface composition of the star. Here we want to study the impact on the surface rotation of the star.

In the first paper of this series \citep{paperI}, we followed the orbital evolution of a planet accounting for all the main effects impacting the orbit (changes of masses of the star and the planet, frictional and gravitational drags, tidal forces) and computed the changes of the stellar rotation due to the planet orbital changes. We used rotating stellar models allowing to follow in a consistent way the angular momentum transport inside the star.
In the present paper, we study what happens when the orbital decay makes the planet to be engulfed by the star.
More precisely we address the following questions:

\begin{itemize}
\item By how much the surface rotation can increase through a planet engulfment process?
\item  How long is the period during which a rapid surface velocity can be observed after an engulfment?
\item  Can the increase of the surface velocity  trigger some internal mixing?
\item  Are there other signatures in addition to fast surface rotation linked to a planet engulfment event?
\end{itemize}

In Section \ref{sec:Ingredients_models}, we discuss the physics included in our models 
Section \ref{sec:impact_s_v} presents post-engulfment evolution of various stellar models. 
Comparisons with observations are discussed in Sect.~\ref{compobs}.
Finally in Sect.~\ref{sec:7} the main results are listed.

\section{Ingredients of the models}\label{sec:Ingredients_models}

\subsection{The stellar models} \label{sec:stellar_models}
The rotating stellar models are computed using the Geneva stellar evolution code \citep[for a detailed description see][]{egg08}.
The reader can refer to \citet{ekstrom12} for all the details about the ingredients of the models (nuclear reaction rates, opacities, mass loss rates, initial composition, overshooting, diffusion coefficients for rotation) and to \citet{paperI} for how the exchanges
of angular momentum between the planetary orbit and the star are treated.

The present models allow to make predictions of the evolution of the surface and also of the  interior velocity resulting from the following  processes: shear turbulence and meridional currents in radiative zones, convection, changes of the structure, loss
of angular momentum by stellar winds, changes of angular momentum of the star resulting from tidal forces between the star and the planet and due to the process of planet engulfment. In convective zones, solid body rotation is assumed, while in radiative zone
radial differential rotation can develop.

\subsection{The planet model}\label{sec:pl_phen}

Some knowledge of the structure of the planet is needed to determine where, once engulfed,  the planet dissolves and deposits
its angular momentum into the star. More precisely, we need to know two quantities (see Eq.~\ref{equa:tvirial} below): the planet radius, $R_{\rm pl}$, and
the mean molecular weight of the planet gas,  $\mu_{\rm pl}$.

The radius of the planet is determined using the mass-radius relation by \citet{zapolsky69}
\begin{equation}
R_{\rm pl}\approx 0.105 \left( \frac{2x^{1/4}}{1+x^{1/2}}\right)\ \ [R_{\odot}]\ \ \ \ ,
\end{equation}
with $x = M_{\rm pl}/(3.2\times 10^{-3})$ and M$_{\rm pl}$ is the planetary mass given in solar mass. 

The same initial chemical composition is considered for the star and the planet. Here we consider a
mass fraction of hydrogen $X = 0.707$, a mass fraction of  helium $Y = 0.273$ and a mass fraction of the heavy elements $Z = 0.02$. With that composition, $\mu_{\rm pl} = 0.614$.
Below, we shall consider different compositions for the planet and the star for what concerns carbon
and lithium. These changes have however very limited impacts on the value of $\mu_{\rm pl}$ and are neglected 
in the estimate of this quantity.

\subsection{Physics of the engulfment}

In \citet{paperI}, given a mass of a star and a mass of the planet, we studied
the range of initial semi-major axis that leads to a planet engulfment along the red giant branch.
Now, in order to describe what will happen next, we need some prescriptions for the fate of the planet inside the star.
According to the work by \cite{soker84}, planets with masses inferior or equal to about 20 M$_J$
will be destroyed in the star. In the present work, we consider only planets with masses equal or below
15 M$_J$, and therefore we assume that they are destroyed.

In order to be able to compute the impact
on the surface velocity of this destruction process we need to know two quantities. The first one is the timescale for the
destruction of the planet and the second one is the location of this destruction inside the star. For instance, if the destruction occurs in a very short timescale (much shorter than the evolutionary timescale) then, once engulfment occurs, the whole orbital angular momentum of the planet orbit will be given to the star in one shot. Moreover, if the destruction occurs in the convective envelope, then this angular momentum can be added in a very
simple way to the convective envelope (see below).

\subsubsection{Destruction timescale}

\citet{livio84} studied the evolution of star-planet systems during the red giant phase. They find that the engulfed planets with masses equal or below 10 M$_J$ are destroyed after a few thousand years (see their Fig.~3). This is very short with respect to the evolutionary timescale during the red giant phase. Indeed, the duration of the first ascent of the red giant branch branch ({\it i.e.}, before the ignition of helium burning in the core) is 180 Myr for a 1.7 M$_\odot$ star.  Therefore, we can consider that the angular momentum that remains in the orbit of the planet at the moment of the engulfment will be delivered to the star in one shot. 

\subsubsection{Where is the angular momentum deposited?}

The location of the planet dissolution ({\it i.e.} the dissolution point)  depends on the physical mechanism that is responsible for it. We can consider two main mechanisms: thermal and mechanical destruction of the planet. Thermal dissolution is obtained where the virial temperature of the planet becomes smaller than the local stellar temperature. Deeper beyond this point, the thermal kinetic energy of the stellar material is larger than the binding energy of the planet. 
 
Knowing the internal structure of the star, it is possible to compare the local temperature in the star and the virial temperature of the planet \citep{siess99I}:
\begin{equation}
T_{\rm v,pl}\sim \frac{GM_{\rm pl}\mu_{\rm pl}m_{\rm H}}{kR_{\rm pl}} \sim 10^5   \mu_{\rm pl} {M_{\rm pl}\over M_J} {R_J \over R_{\rm pl}} [K]\ \ \ \ .
\label{equa:tvirial}
\end{equation}
where $k$ is the Boltzmann constant,  $M_J$, $R_J$, the mass and radius of Jupiter.

\begin{figure*}
\includegraphics[width=.49\textwidth, angle=0]{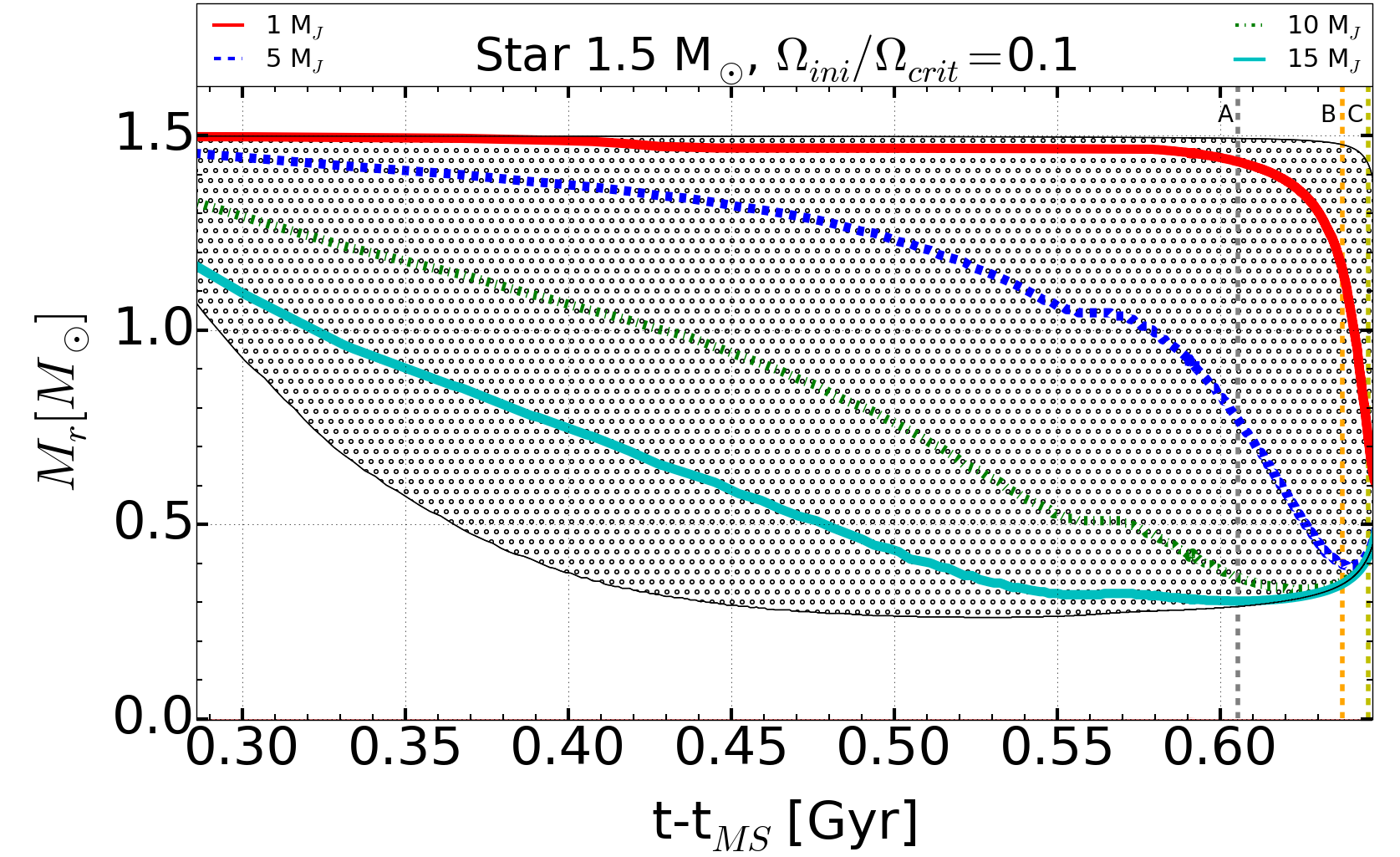}\includegraphics[width=.49\textwidth, angle=0]{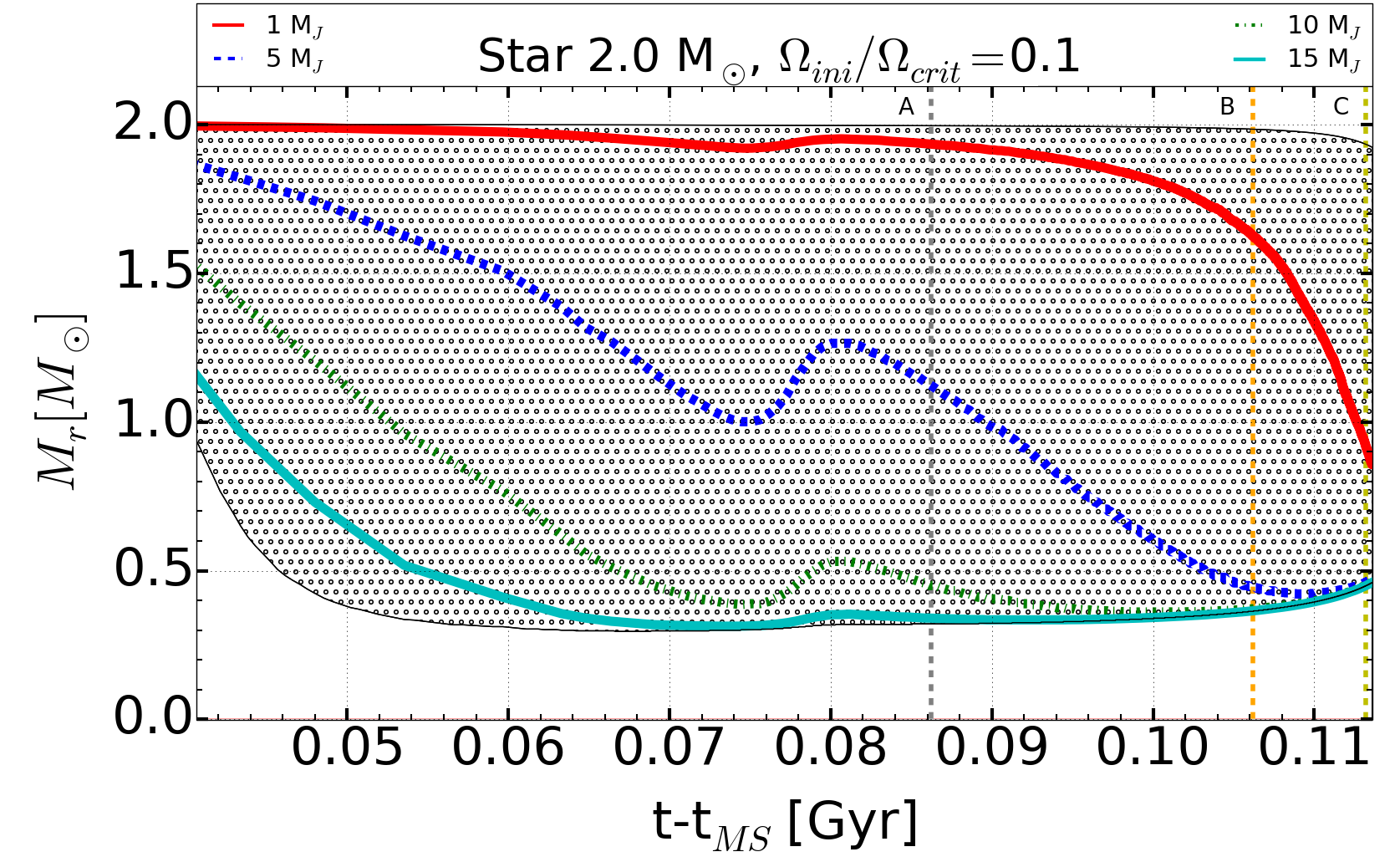}
\caption{Kippenhahn diagram during the red giant phase. The central radiative zone and the convective envelope are represented respectively by a white area and an area with black dots.
The horizontal axis is the age of the star minus its Main-Sequence lifetime (t$_{\rm MS}$).
The four lines indicate the limit layers that can be reached by the planets of mass 1 M$_{\rm J}$ (red solid line), 5 M$_{\rm J}$ (blue dashed line), 10 M$_{\rm J}$ (green dash-dot line) and 15 M$_{\rm J}$ (continuous cyan line).
The vertical dashed-lines labeled A, B and C indicate when the engulfment occurs for respectively the 15 M$_J$ planet at an initial distance of 0.5 au, the 10 M$_J$ planet at an initial distance of 1.0 au and the 1 M$_J$ planet at an initial distance of 1.5 au.
All the other cases occur between the lines A and C. 
The left panel corresponds to a model star of 1.5 M$_{\odot}$, the right panel to a mass of 2 M$_{\odot}$.   
}
\label{fig:where15}
\end{figure*}

In Fig.~\ref{fig:where15}, the regions where the stellar temperature is equal to the virial temperature of planets of various masses  are indicated. The lines indicated in Fig.~\ref{fig:where15} are isothermal lines\footnote{These lines would not be isothermal lines in case the structure of the planet would change during its journey inside the star. Here it is assumed that during the very short migration time
the planet structure does not change significantly. The only change occurs at the very end, when the planet dissolves.}.
We also have indicated by vertical dashed lines the engulfment time of planets of various masses and initial semi-major axis as obtained in \citet{paperI}.
As indicated in Sec.  2.3.1, the migration time of the planet is very short, thus the destruction of those planets will occur at the intersection
of the isothermal line for the planet mass considered with the corresponding vertical lines indicating the time of engulfment (only a few cases are shown for illustration). A few interesting points can be noted looking at Fig.~\ref{fig:where15}:
\begin{itemize}
\item When the star evolves, the dissolution points reach (in general) deeper layers in the stellar interior. This is a consequence of the expansion of the envelope during the red giant phase that produces a lowering of the temperature at a given Lagrangian mass. Only when the star contracts, for example during the bump, the dissolution point shifts for a time outwards. This explains the local minimum that can be seen for instance for the 5 M$_J$ mass planet at a time around 0.08 Gyr for the 2 M$_\odot$ model (right panel of Fig.~\ref{fig:where15}).
\item More massive planets go deeper inside the star. This is of course expected, since a more massive planet requires more extreme conditions to be destroyed than lighter ones. 
\item Under the assumptions above and in the light of Fig.~\ref{fig:where15}, it is reasonable to assume that the planets are destroyed in the convective envelope of the star.
\end{itemize}

\cite{siess99I} also examined the possibility that when the planet gets closer to the stellar core, tidal effects induce strong distortions of the planet that can destroy it. Using the elongation stress at the centre of the planet as approximated by \citet{soker87}, we find in most cases that the planet would be destroyed in the convective envelope as is the case when the criterion based on the virial temperature is used. Only in a few cases, the planet would be destroyed by tides just below the convective envelope. In the following we shall consider only the criterion based on the virial temperature and thus assume that all our engulfed planets will deliver their  angular momentum in the convective envelope.

Let us call $\mathcal{L}_{\rm pl}$ the angular momentum of the planet orbit at the time of engulfment and $\mathcal{L}_{\star}(ce)$ the angular momentum of the external convective envelope ($ce$) of the star just before
engulfment. To obtain the new angular velocity of the envelope after engulfment, $\overline{\Omega}$, we write
\begin{equation}
\overline{\Omega}(ce) = \Omega(ce)\left(1+\frac{\mathcal{L}_{\rm pl}}{\mathcal{L}_{\star}(ce)}\right)\ \ \ \ ,
\end{equation}
where $\Omega (ce)$ is the angular velocity of the convective envelope just before the engulfment. 

 Equation (3) assumes that the moment of inertia of the convective envelope is not changed by the planet engulfment process. Actually it might happen that the engulfment process would
add thermal energy in the upper layers of the stellar envelope, making these layers to expand for a while. But the excess of energy will be rapidly radiated away and thus the star will evolve back rapidly to its
initial state. Just as a numerical example, in case all the kinetic energy in the planetary orbit of a 15 M$_J$ initially orbiting a 1.7 M$_\odot$ at 0.5 au would be added as an increase of the internal energy in the outer layers, 
then the excess of energy would be radiated away in about 55 years, so in a very short time compared to the evolutionary timescale.

\subsection{Initial conditions considered}

We consider stars with initial masses in the range between 1.5 and 2.5 M$_{\odot}$, with a metallicity $Z = 0.02$ and an initial rotation
equal to $\Omega_{\rm ini}/\Omega_{\rm crit} = 0.1$ and 0.5, where $\Omega_{\rm ini}$ is the initial angular velocity and $\Omega_{\rm crit}$
the critical angular velocity on the ZAMS. These correspond to initial surface velocities  of $\sim$30 and 160 km s$^{-1}$ respectively.

 As explained in paper I, we focused here on stellar masses larger or equal to 1.5 M$_\odot$, because stars in this mass range 
do not have a sufficiently extended outer convective zone to activate a dynamo during the main sequence, so that unless they host a fossile magnetic field, they do not undergo any significant surface magnetic braking. The results obtained in the present work are thus not sensitive to the uncertainties related to the modeling of this braking during the main sequence. In contrast, lower initial mass stars have an extended outer convective zone during the main sequence. Therefore they can activate a dynamo and suffer a strong magnetic braking. The evolution of the rotational properties of these lower initial mass stars is then different and will be the topic of another paper in this series.

Planets with masses equal to 1, 5, 10 and 15 M$_{\rm J}$ have been considered\footnote{In our models, we take planets with masses not larger than 15 M$_{\rm J}$ that means below the limiting value
 $M_{\rm cri} \simeq$ 20 M$_{\rm J}$ estimated by \citet{soker84}, above which the planet is no longer completely destroyed when engulfed by the stellar envelope.
 This limit is also consistent with the value found in the numerical simulations by \citet{Staff2016}.}. 
The initial semi-major axes ($a_{\rm 0}$) have been taken equal to 0.5, 1 and 1.5 au and the eccentricities of the orbits are equal to 0. 

In \citet{paperI}, the evolution of the planetary orbit and of the star have been computed in a consistent way up to the point of engulfment. 

The end of the computation in \citet{paperI} represents the initial conditions for the present work (see Table~\ref{table:stru0.1} in the \hyperref[sec:appe]{Appendix}).

\section{Impact of planet engulfment on the surface velocities of red giants}\label{sec:impact_s_v}

The impact of tides and engulfment on the surface velocity of a red giant depends at least on the following parameters: the mass of the planet, its initial distance to the star and the mass of the star. Other parameters like the metallicity (not discussed here) and the rotation
(that will be discussed here) of the star also affect the results. 
\begin{figure*}
 \includegraphics[width=.49\textwidth, angle=0]{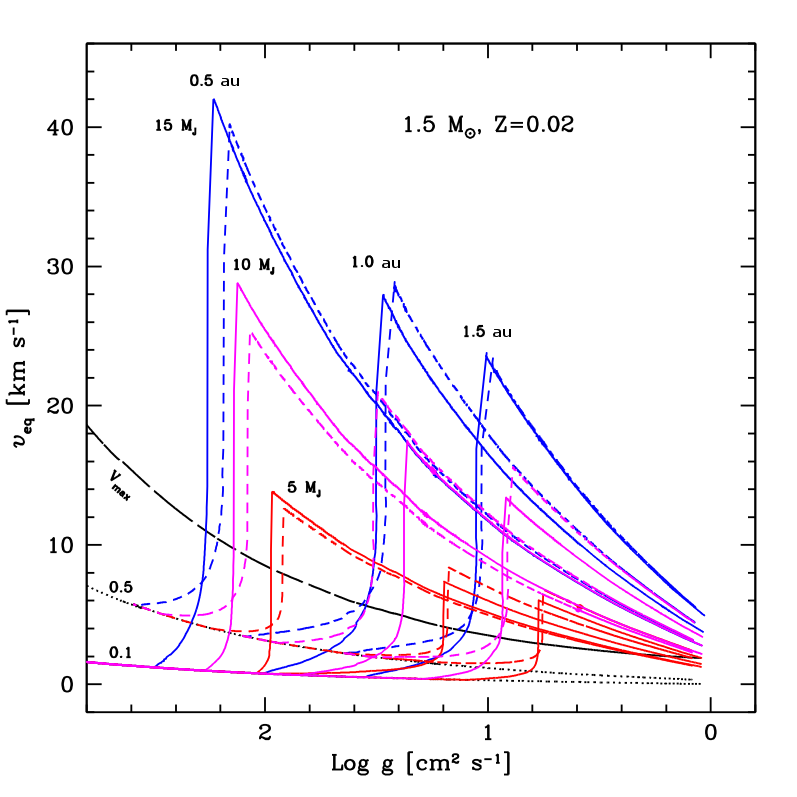}\includegraphics[width=.49\textwidth, angle=0]{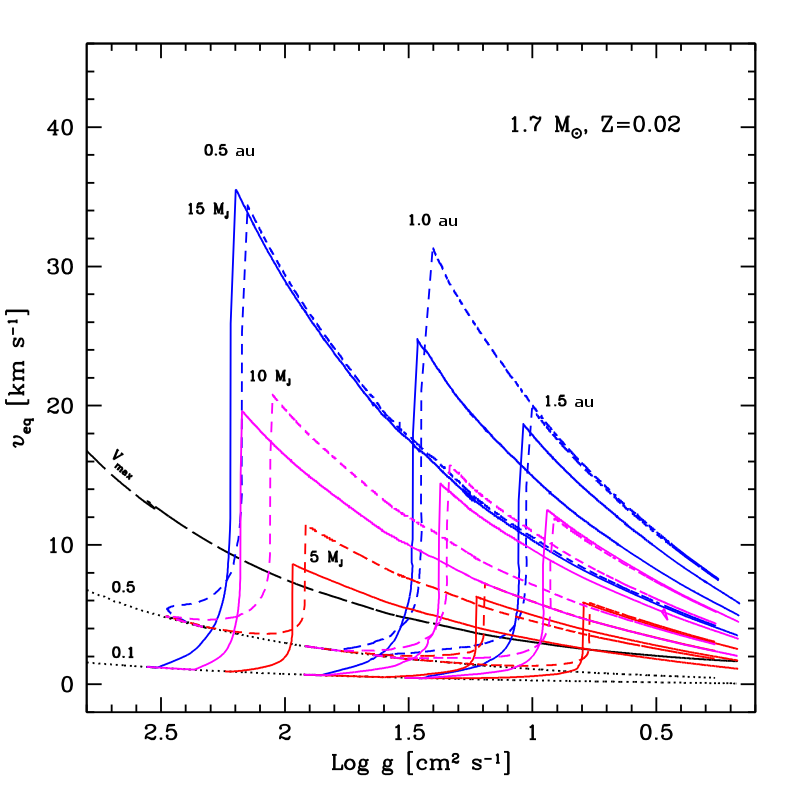}
 \includegraphics[width=.49\textwidth, angle=0]{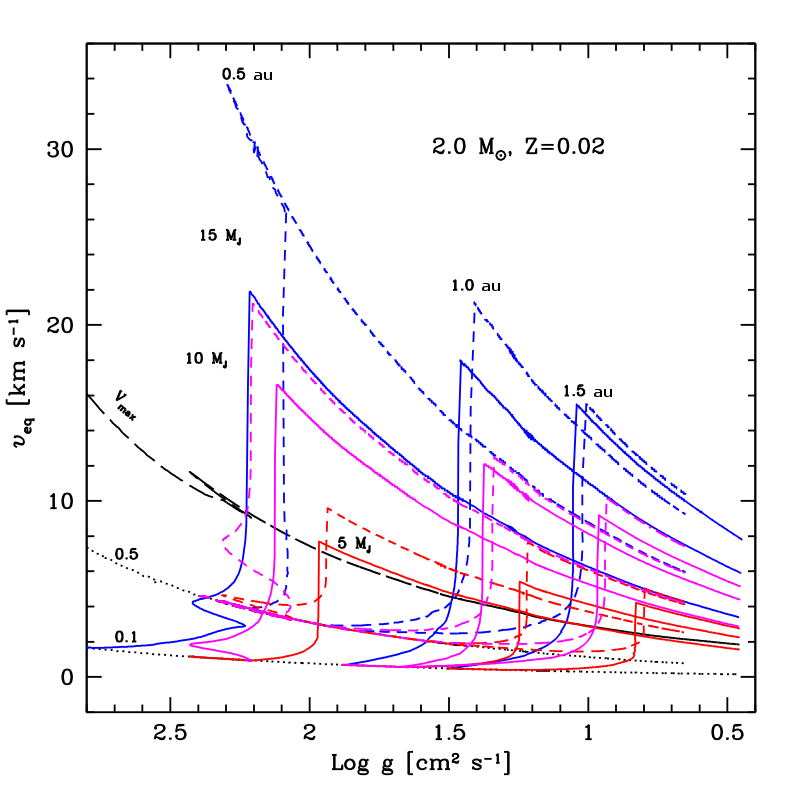}\includegraphics[width=.49\textwidth, angle=0]{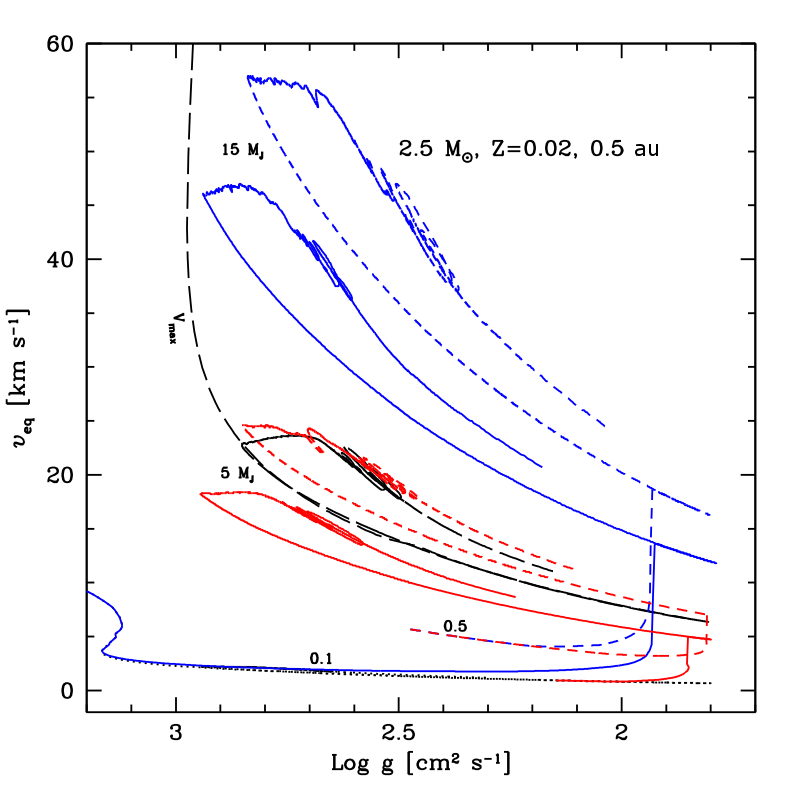}
\caption{Evolution of the surface velocity as a function of the logarithm of the surface gravity. Dotted (black) lines are for stars without engulfment. Cases
for $\Omega/\Omega_{\rm ini}=0.1$ and 0.5 are shown. The black long-dashed line (labeled V$_{\rm max}$) corresponds to solid body rotation without engulfment starting with an initial rotation near the critical velocity (see text). 
Continuous lines correspond to cases with planet engulfment for stars with an initial velocity equal to 10\% the critical one and the short-dashed lines correspond to the same cases but for an initial rotation equal to 50\% the critical one. 
The different colors for solid and dashed lines are for planet masses of 5 (red), 10 (magenta), and 15 M$_J$ (blue)  and initial orbital distances of 0.5, 1, and 1.5 au from left to right (see labels).
The different panels are for different values of the mass of the star. In the case of the 2.5 M$_\odot$, the evolution is pursued until the end of the core He-burning phase, while for lower initial mass stars, the evolution is stopped at the tip of the red giant branch.}
\label{fig:vlog}
\end{figure*} 

\begin{figure*}
\includegraphics[width=.49\textwidth, angle=0]{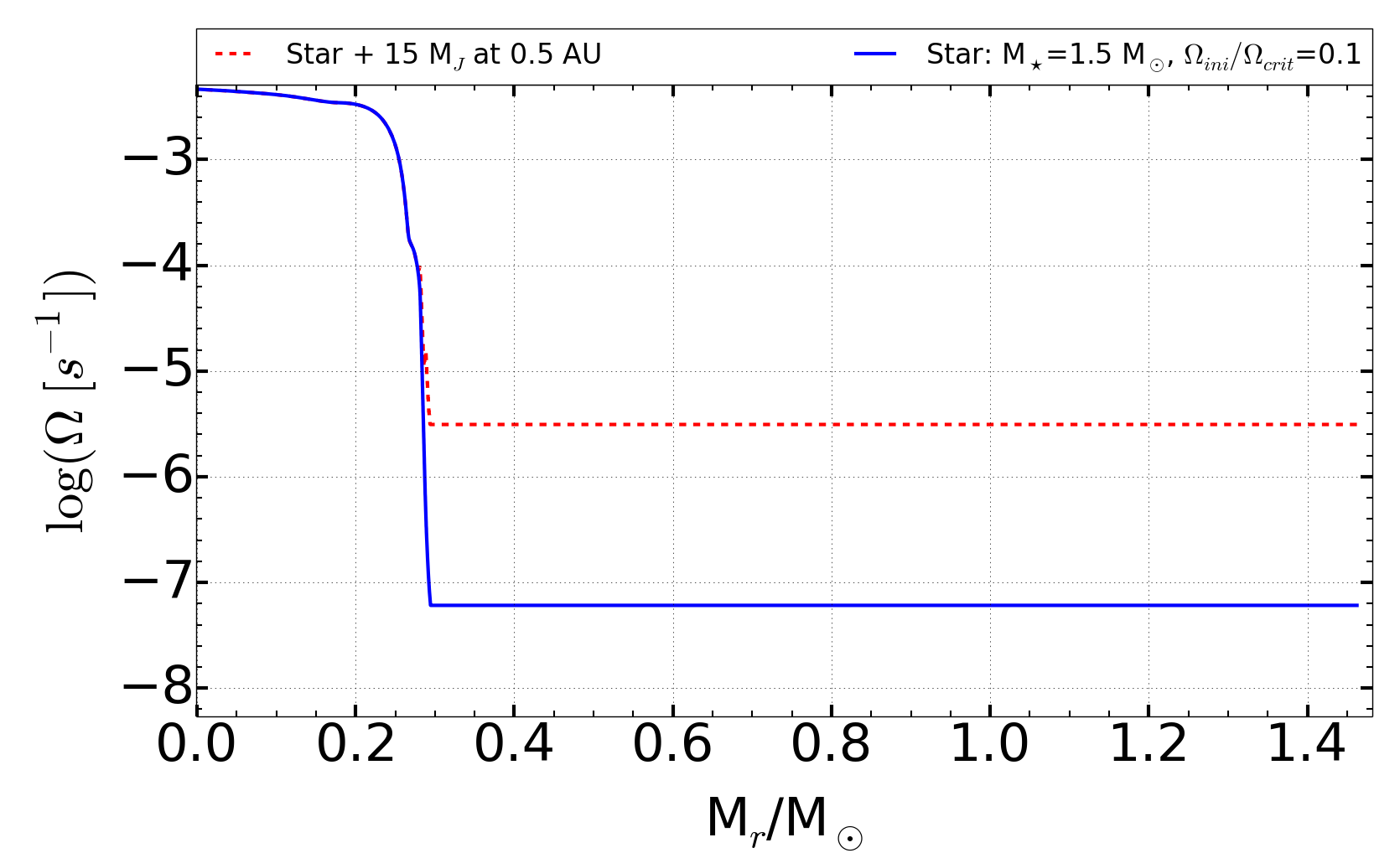}\includegraphics[width=.49\textwidth, angle=0]{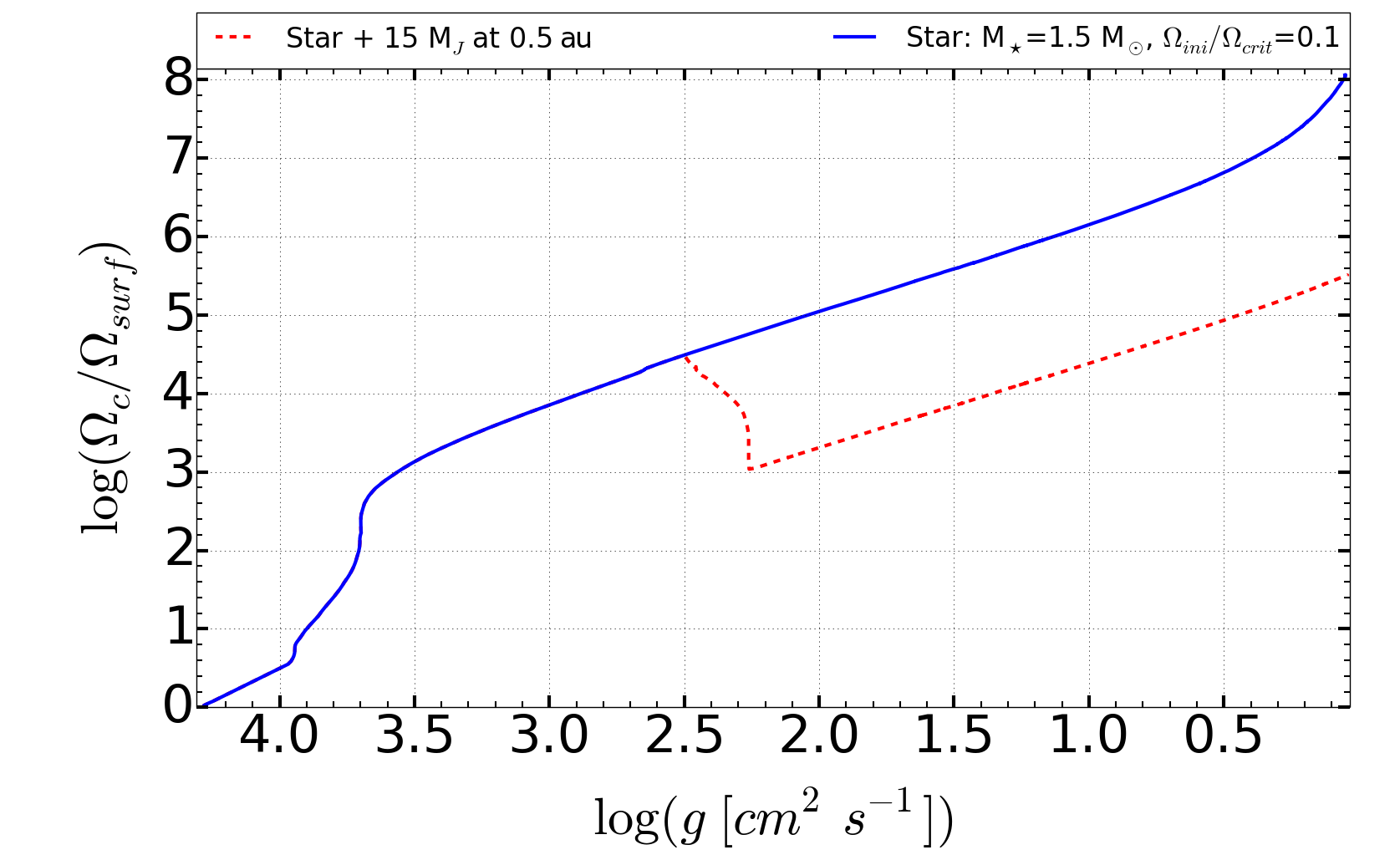}
\caption{{\it Left panel:} Variation of the angular velocity as a function of the Lagrangian mass inside a 1.5 M$_\odot$ model along the red giant branch just before (blue continuous line) and after the engulfment (red dashed line) of a 15 M$_J$ mass planet at an initial distance of 0.5 au from the star. 
{\it Right panel:} Evolution of the ratio between the angular velocity of the core and the surface angular velocity as a function
of the surface gravity along the red giant branch for the same models as in the left panel.}
\label{omega}
\end{figure*}

Figure~\ref{fig:vlog} shows the evolution of the surface equatorial velocities for the present stellar models with and without engulfment of a planet. Tables \ref{table:Omega0.1} and \ref{table:Omega0.5} in the \hyperref[sec:appe]{Appendix} present some characteristics of the star after an engulfment.

\subsection{Planet engulfment by a 1.5 M$_\odot$ star}

Let us begin by discussing the case of the 1.5 M$_\odot$ stellar model (upper left diagram in Fig.~\ref{fig:vlog}).
 The case of an engulfment of a 15 M$_J$ initially at a distance of 0.5 au is considered for the purpose of the discussion.
We can note the following features:
\begin{itemize}
\item Planet engulfment may have a very strong effect on the surface rotation of a red giant. Compare for instance in the upper left panel, the (black) dotted line that shows the evolution of the surface velocity of a 1.5 M$_\odot$ red giant having begun its evolution with an angular velocity equal to 50\% the critical velocity (142 km s$^{-1}$) and with no planet engulfment, with the (blue) dashed line showing the evolution of the surface velocity of a similar star engulfing a planet.
We see that while the isolated star has a surface velocity that remains below 7 km s$^{-1}$ for log $g$ below 2.8, the star having engulfed the planet shows an extremely rapid increase of the surface velocity up to a maximum value of about 40 km s$^{-1}$. 
\item The left panel of Fig.~\ref{omega} shows the variation of the angular velocity inside the 1.5 M$_\odot$ with an initial angular velocity equal to 10\% the critical velocity just before and just after the engulfment of a planet. The angular velocity in the convective zone extending from a mass equal to about 0.3 M$_\odot$ up to the surface has its angular velocity increased by about 1.7 dex (a factor 50). 

At engulfment (around a log $g$ equal to 2.5), the  ratio between the angular velocity of the core and that of the envelope
decreases from a value of about 32 000 to a value equal to 1000 (see the right panel of Fig.~\ref{omega}) . Then the contrast increases again as a result of the expansion of the envelope and the contraction of the core.
\item The present models cannot account for the small contrast of about one dex between the core and the envelope rotation rates that has been obtained by asteroseismology for red giants at the base of the giant branch \citep{beck12,Mosser2012,deh12,deh14}. In that respect, an additional internal angular momentum transport process is missing in the radiative interior of the present models \citep{egg12,mar13,can14}. This weakness has however no major impact on the process studied here. The missing transport mechanism would add a fraction of the angular momentum of the core to the envelope, but this will not much accelerate the envelope because the angular momentum in the core is quite small with respect to the one contained
in the envelope. Therefore the angular momentum coming from the planetary orbit remains in any case much larger than the angular momentum of the envelope and the final result of a planet engulfment would remain nearly unchanged.
\item After the engulfment, the surface velocity decreases because the stellar envelope continues to expand but it remains above 8 km s$^{-1}$ for a time which is nearly 36 Myr (see Table~\ref{table:Omega0.5}) that means for about 15\% of the total duration of the red giant branch phase. No significant changes of the structure of the star (change of the extent of the convective envelope and/or of the mixing of the chemical elements) is produced by the increase of the rotational speed of the convective envelope. So in our models, any changes of the surface abundances can only be attributed to the dilution of the planet material into the convective envelope. This will be briefly investigated in the next section when comparisons with observations are discussed. 

\item In Fig.~\ref{fig:vlog} we have also indicated the maximum value for the surface rotation that is expected during the red giant phase (see the black long-dashed line) for a star with no planet. This line has been obtained assuming that the star rotates near the critical velocity on the ZAMS and that a solid body rotation is maintained at every time.
This is clearly an extremum. This case of solid body represents an extremum case in terms of efficiency of angular momentum transport
and also in terms of surface velocities\footnote{One could argue that, once the solid body rotation is reached,  some processes might still advect angular momentum from the center to the surface producing a negative gradient of $\Omega$, {\it i.e.} a core rotating slower than the envelope, and increasing the surface velocity. However,
this is not supported by asteroseismic determinations of the interior rotation of red giants \citep[e.g.][]{beck12,Mosser2012,deh12,deh14} that show that red giants have a clear increase of rotation towards the central regions. In that respect, the hypothesis of solid body rotation gives
a very conservative upper limit for the surface rotation of red giants.}. This means that,
any observed star, with initial masses of about 1.5 M$_\odot$, with a surface velocity larger than the one given by the curve $V_{\rm max}$ in the left upper panel of Fig.~\ref{fig:vlog} needs some acceleration process which might come either from tidal forces or from an engulfment.

We see that
for a 1.5 M$_\odot$ star at a metallicity Z=0.02, this maximum velocity is well below the 40 km s$^{-1}$ reached by the engulfment of a 15 M$_J$ planet with an initial distance to the star equal to 0.5 au. {\it Therefore planet engulfment may be a sufficiently strong mechanism to produce surface acceleration beyond any reasonable process that can occur in single stars}. Of course not any planet engulfment will produce such strong acceleration, but at least some events can produce surface velocities beyond what can be reasonably explained by single star evolution.
\item We see that decreasing the initial distance between the star and the planet produces higher surface velocities and longer periods during which high surface rotations can be observed. The main reason for this is that smaller the initial distance, earlier along the red giant branch the engulfment occurs. Earlier the engulfment, smaller is the moment of inertia of the convective envelope and thus more important will be the effect produced by the injection of the planetary angular momentum in the stellar convective envelope. We could argue however that larger the distance, larger the amount of angular momentum in the planet orbit. However the effect indicated above (the small moment of inertia of the envelope) is the most important one and clearly overcomes the effect of the larger angular momentum associated to a larger orbital radius. An interesting consequence of that point was noted already by \citet{carlberg09, carlberg11}: planet engulfment produces the largest increases of the surface rotation when the engulfment occurs at the beginning of the red giant branch. 
\item We see also that the higher the mass of the planet, the larger the maximum surface velocity reached (assuming identical initial distance of the planet to the star) and longer the period during which the surface velocity is superior to a given limit (see Tables \ref{table:Omega0.1} and \ref{table:Omega0.5}).
\item If keeping all the other parameters equal, we change the initial rotation of the star, passing from $\Omega_{\rm ini}/\Omega_{\rm crit} = 0.1$ to 0.5, we obtain very similar behaviors. This indicates that the outcome at least in the case of the 1.5 M$_\odot$ does not much depend
on the initial rotation of the star.
\end{itemize}

\subsection{Planet engulfment by stars with masses between 1.7 and 2.5 M$_\odot$}

\begin{figure}
\includegraphics[width=.49\textwidth, angle=0]{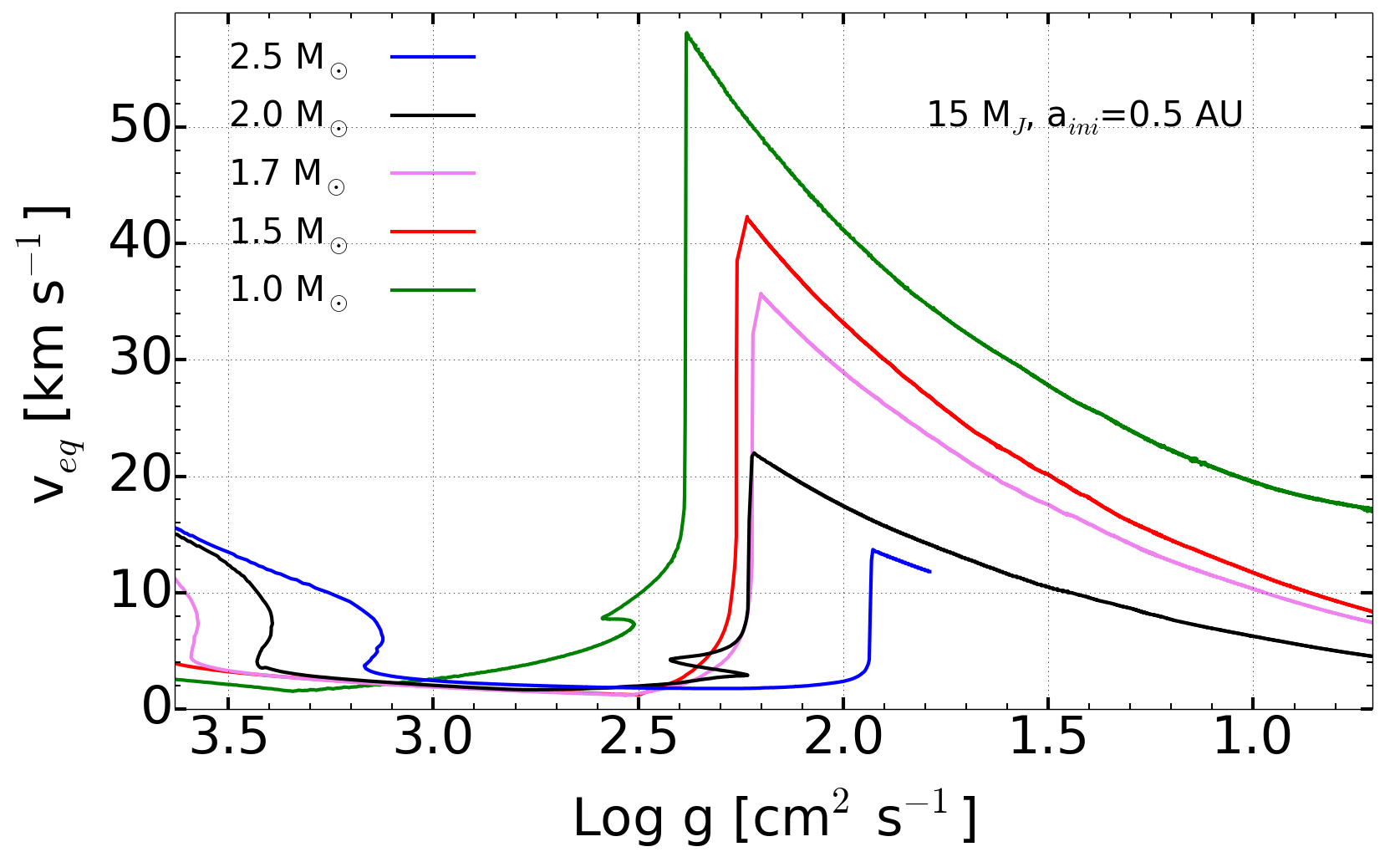}
\caption{Evolution of the surface velocity of various initial mass stellar models during the red giant phase as a function of the surface gravity. 
The initial rotations are between 30 and 50  km s$^{-1}$.
Planets have a mass equal to 15 M$_J$ and the orbits have an initial semi-major axis equal  to 0.5 au.}
\label{masse}
\end{figure} 

  In Fig.~\ref{masse}, the increase of the surface velocity due to a planet engulfment is compared for stars with different initial masses. We see that in general, for a given planet mass and a given initial distance between the star and the planet, the engulfment occurs at a higher surface gravity when the mass of the star decreases. The maximum surface velocities that are reached are higher for smaller initial mass stars. We have added a track for a one solar mass model to provide a clear illustration of this. As explained in Sect.~\ref{sec:Ingredients_models}, such a star suffers magnetic braking during the Main-Sequence phase. This is introduced for the one solar mass model by using a solar-calibrated value for the efficiency of this braking. As a result, the surface velocity at the base of the red-giant branch is much lower for the one solar mass model than for stars with higher initial masses. When the engulfment occurs, the acceleration is larger for this model than for the more massive stars, because of the larger ratio between the planetary orbital angular momentum and the stellar envelope angular momentum.
Since the 2.5 M$_\odot$ star does not evolve through the He-flash, we pursued the evolution during the core He-burning phase. It is why its evolution looks so different from the lower initial mass stars  (compare the lower right panel of Fig.~\ref{fig:vlog} with the other panels). Actually, just after
the engulfment, we can see a very similar behavior as those we could see in the other panels. The surface velocity increases nearly vertically at engulfment and after engulfment decreases with the surface gravity. When the tip of the red giant branch is reached, the star contracts
and the track evolves again at high surface gravities following first the same path as the one used when the star was evolving up the red giant branch, passing again through the same point where engulfment occurred and evolving still to higher surface velocities when it
continues to contract. Most of the core He-burning phase will occur along the extremity of the loop at surface gravities between 2.7 and 3.0.
After the core He-burning phase, the envelope of the star expands, the surface gravity decreases and the star join the asymptotic giant branch.

Interestingly, the surface velocity during the core He-burning phase remains quite high after an engulfment (above the curve $V_{\rm max}$ in the lower right panel of Fig.~\ref{fig:vlog}). Therefore, the signature of planet engulfment may remain visible for the whole
core He-burning phase. This would likely be the case for lower initial mass stars too evolving through a He-flash although this 
remains to be confirmed by computations. 

In general, as was the case for the 1.5 M$_\odot$ star, changing the initial rotation rate of the star has  not a very large impact on the results. We note however the very thin loop at engulfment in the lower left panel of Fig.~\ref{fig:vlog}
in the case of 2 M$_\odot$ models with an initial rotation equal to 50\% the critical angular velocity and with a 15 M$_J$ at initial distance of 0.5 au. For this model,  the time of engulfment coincides with the time when
the bump occurs. During the bump, the star contracts (hence the increase of the surface gravity seen on Fig.~\ref{fig:vlog}). This contraction occurs when the H-shell enters the domain whose composition has been enriched in hydrogen by the downward extension of the external convective zone. As a consequence, the H-burning shell expands and reduces its temperature. By a mirror effect, the envelope contracts provoking an additional increase of the surface velocity and the thin loop that appears at the end of the engulfment process.

\subsection{How long does a red giant star remain fast rotating after an engulfment?}

\begin{figure*}
\includegraphics[width=.49\textwidth, angle=0]{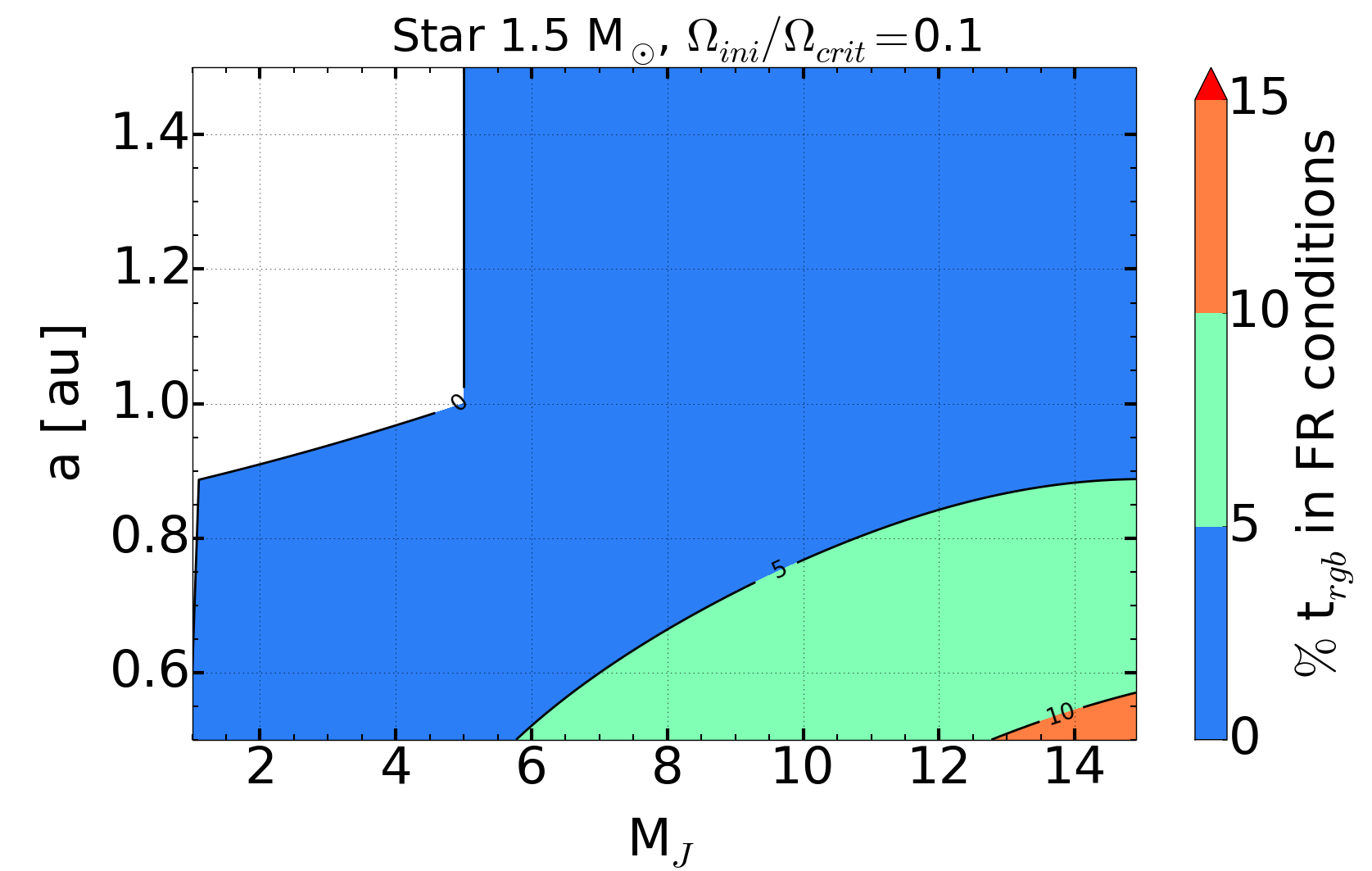}\includegraphics[width=.49\textwidth, angle=0]{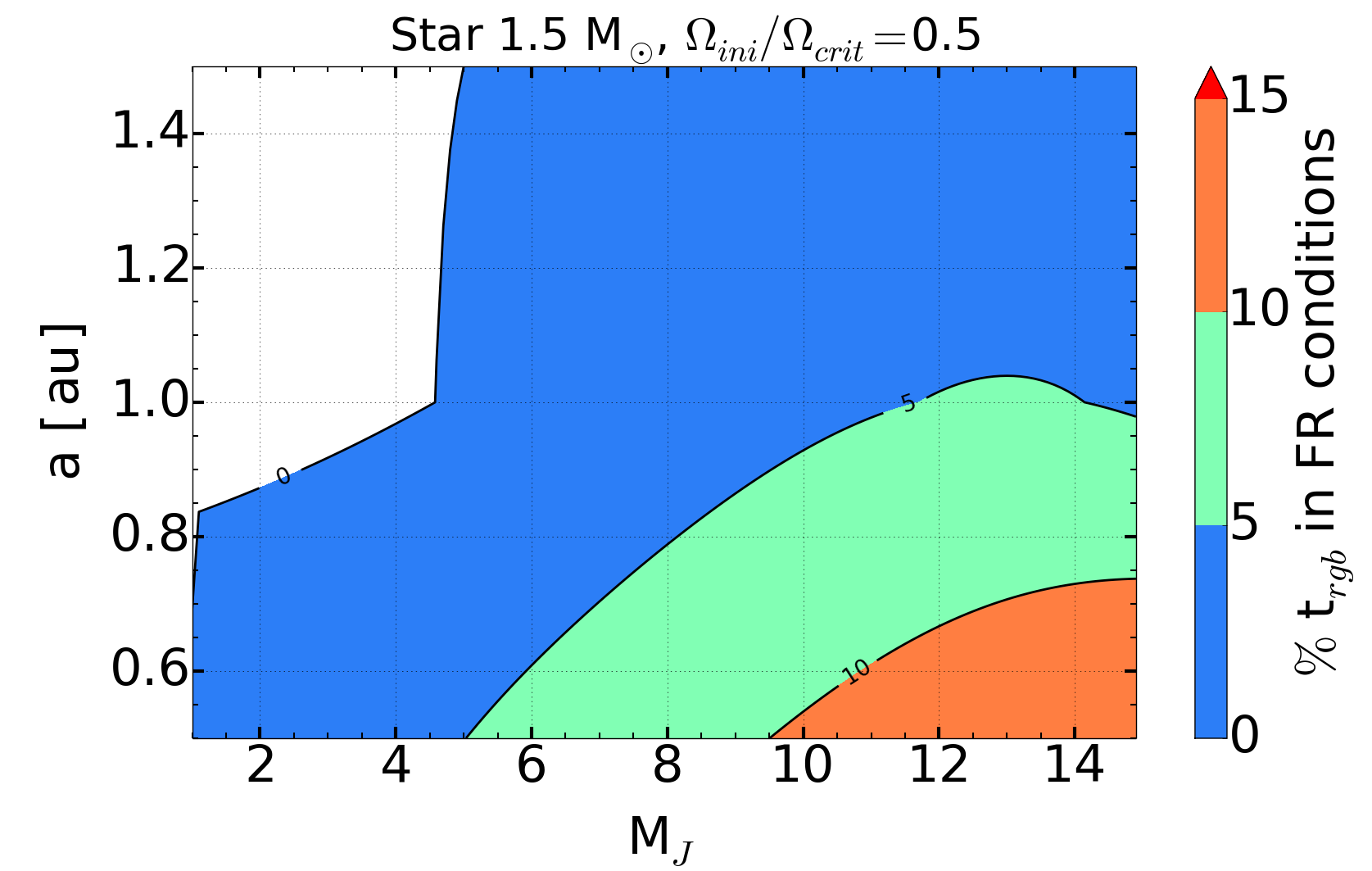}
\includegraphics[width=.49\textwidth, angle=0]{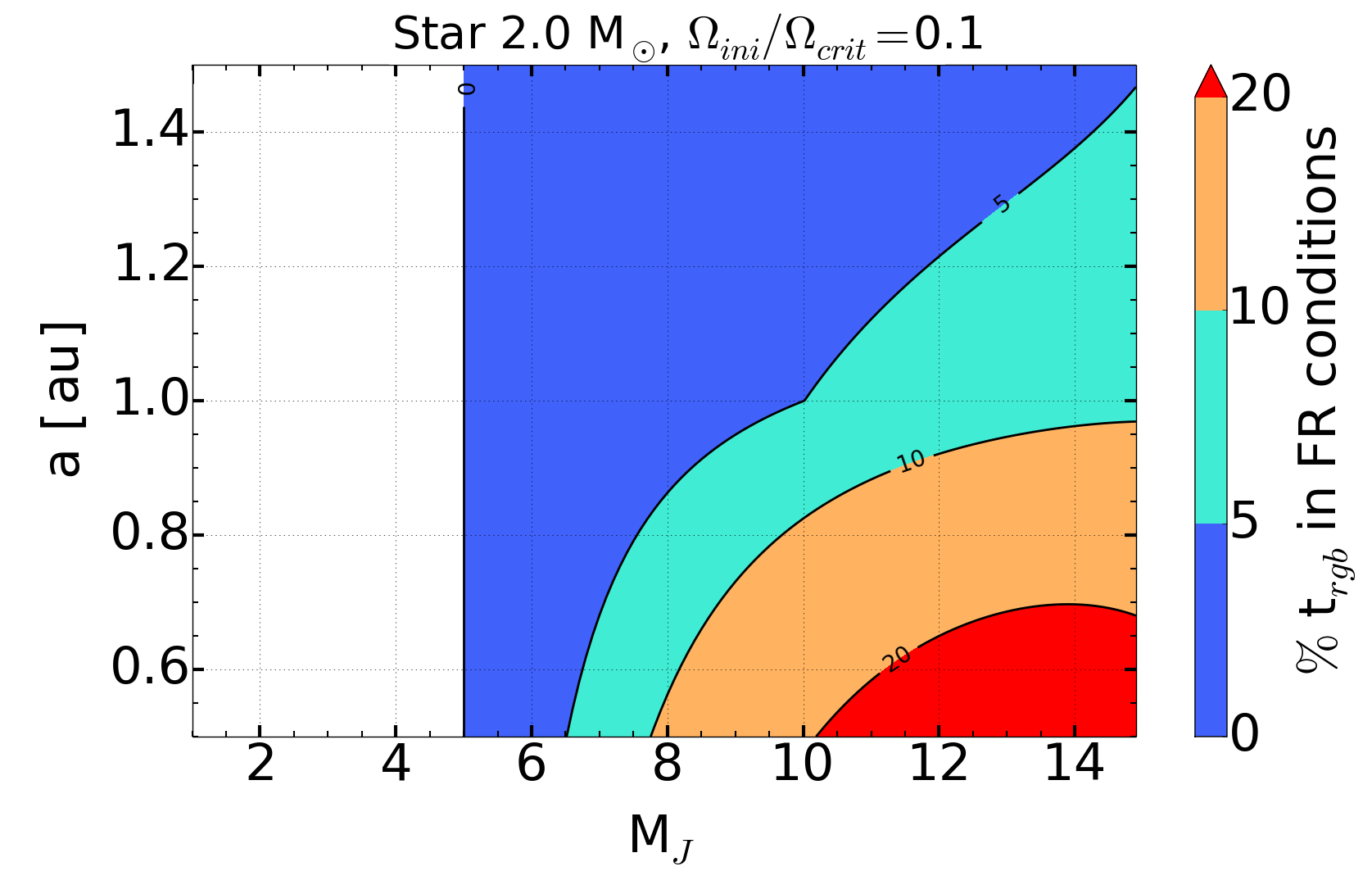}\includegraphics[width=.49\textwidth, angle=0]{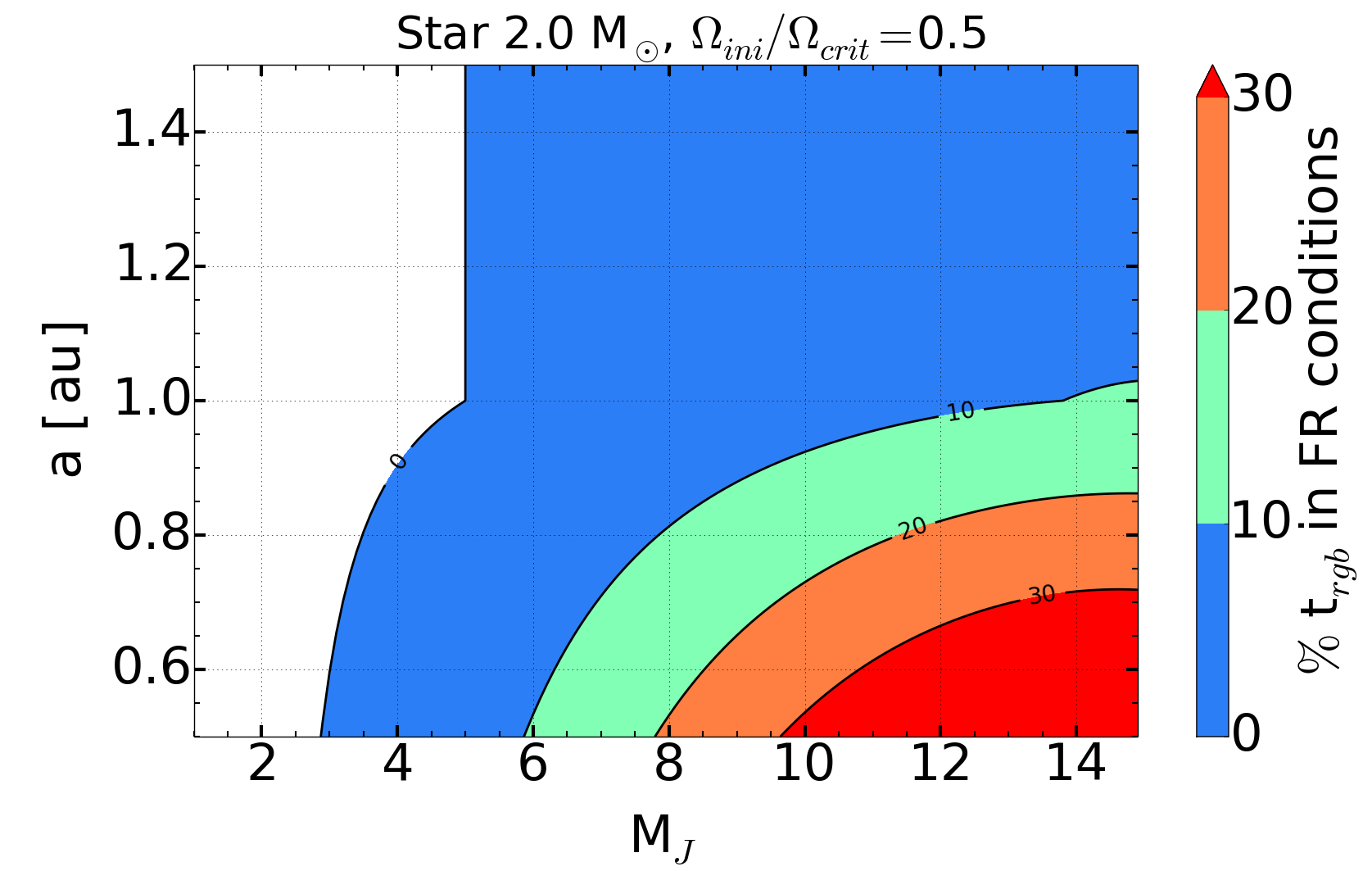}
\caption{Fraction of the total red giant branch duration spent with a surface equatorial rotation velocity larger than 8 km s$^{-1}$ after an engulfment. 
{\it Upper left panel:} Case of a 1.5 M$ _{\odot}$ with an initial rotation $\Omega_{\rm ini}$/ $\Omega_{\rm crit}$ = 0.1.
{\it Upper right panel:} same as the upper left panel, but with $\Omega_{\rm ini}$/ $\Omega_{\rm crit}$ = 0.5.
{\it Lower left panel:} same as the upper left panel, but for a star of mass 2.0 M$ _{\odot}$. 
{\it Lower right panel:} same as the lower left panel, but with $\Omega_{\rm ini}$/ $\Omega_{\rm crit}$ = 0.5.}
\label{fig:percent_fr}
\end{figure*}

In the literature, a value of $\upsilon \sin i$ larger than 8 km s$^{-1}$ is sometimes used as a criterion to qualify a red giant as fast rotating \citep[see {\it e.g}][]{carlberg12}.
We see on Fig.~\ref{fig:vlog} that indeed higher values than 8 km s$^{-1}$ would definitively require some interaction with a second body for surface gravities smaller than Log $g$ equal 2.0 (1.5 M$_\odot$), 2.1 (1.7 M$_\odot$), 2.15 (2 M$_\odot$) and
2.1 (2.5 M$_\odot$). Of course, this does not exclude that stars showing surface velocities below this limit have not undergone any interaction, but in that case, alternative single star models could be proposed to explain the high velocity as well.

Figure \ref{fig:percent_fr} allows to have a synthetic view of the durations of the high surface velocity periods. 
We see that, for a given stellar model, the duration increases when the initial semi-major axis decreases and the mass of the planet increases (see also Tables \ref{table:Omega0.1} and \ref{table:Omega0.5}). 
The conditions for, let us say, having a high surface velocity during about 10\% of the red giant phase are less restricted when the mass of the star passes from 1.5 to 2 M$_\odot$\footnote{The duration of the period during which the surface velocity remains superior to a given limit after an engulfment is 
larger in smaller mass stars. However, when this period is normalized to the total duration of the red giant branch, then the fraction
spent above a given limit decreases when the mass decreases.}. The same occurs when,
keeping the stellar mass constant, the initial rotation passes from 10 to 50\% the critical velocity although the impact is less important than changing the initial mass of the star.

\subsection{Impact of tidal forces with respect to engulfment}

\begin{figure*}
\includegraphics[width=.49\textwidth, angle=0]{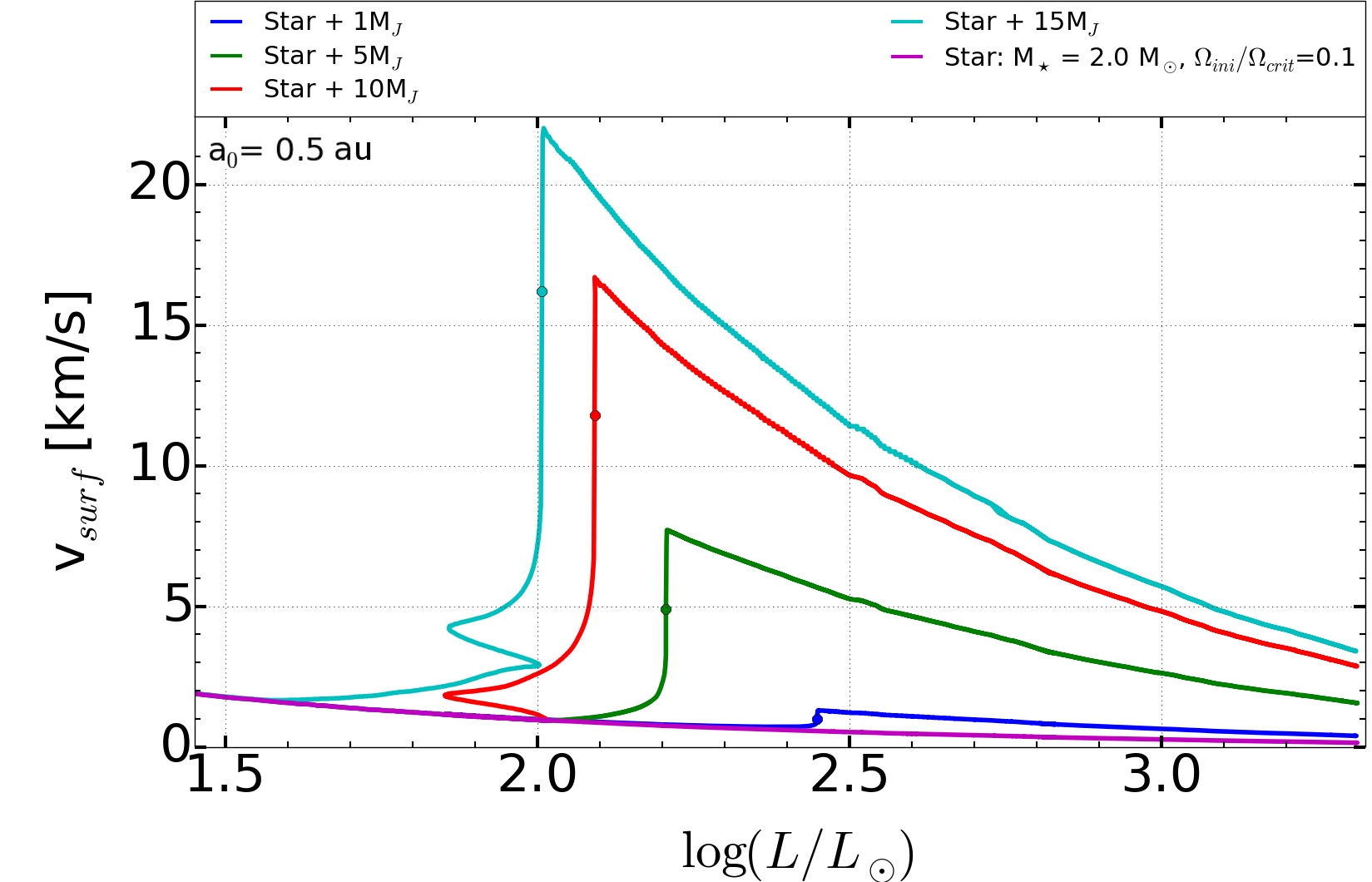}\includegraphics[width=.49\textwidth, angle=0]{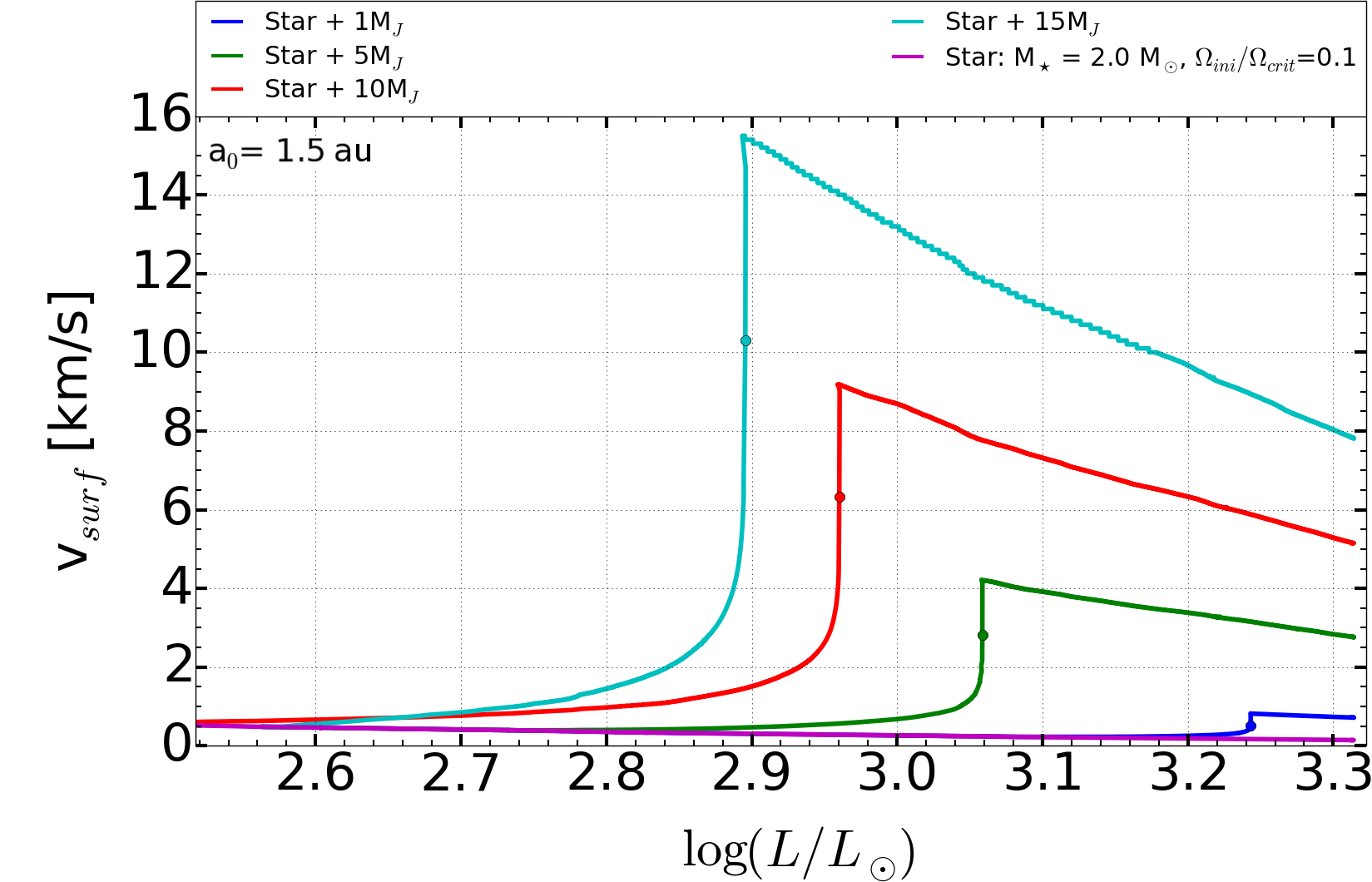}
\caption{Evolution of the surface velocity of a 2 M$_\odot$ stellar model during the red giant phase as a function of the luminosity. The initial rotation of the star is 10\% the critical velocity (on the ZAMS).
When evolution proceeds, the luminosity in general increases except at the bump visible in the left panel under the form of a small hook.
Magenta lines indicate cases without planet engulfment, while blue, green, red and cyan lines correspond to the engulfment of a planet with a mass equal to 1, 5, 10 and 15 $M_{\rm J}$, respectively. The surface velocity reached just before engulfment, {\it i.e.} resulting only from tidal forces, is indicated by a small dot.
{\it Left panel:} The initial semi-major axis is equal to 0.5 au. {\it Right panel:} The initial semi-major axis is equal to 1.5 au.}
\label{fig:vel_surface}
\end{figure*}

The physics of what happens during the engulfment is not so well known and thus we may wonder what is the reliability of the present results.
An interesting point to mention in that context, is the fact that already the tidal forces may lead to significant accelerations of the stellar surface.
Fig.~\ref{fig:vel_surface} shows the evolution of the surface rotation as a function of the luminosity for a few selected models of 2 M$_\odot$ stars.
The small dots along the curves with engulfment indicate the surface velocity reached just before engulfment. The acceleration
up to this point is due only to tidal forces. The acceleration after that point is due to the engulfment process itself.

We see that tides alone may be sufficient to accelerate significantly the surface rotation of the star. 
Since the acceleration due to tides is very rapid \citep[see][]{paperI}, as is the acceleration due to the engulfment, there is no chance using the rotational velocity of the star to distinguish observationally between tides- and engulfment-acceleration processes, would both occur.

\section{Comparisons with observations}\label{sec:compa_ob}
\label{compobs}

\begin{figure}
\includegraphics[width=.49\textwidth, angle=0]{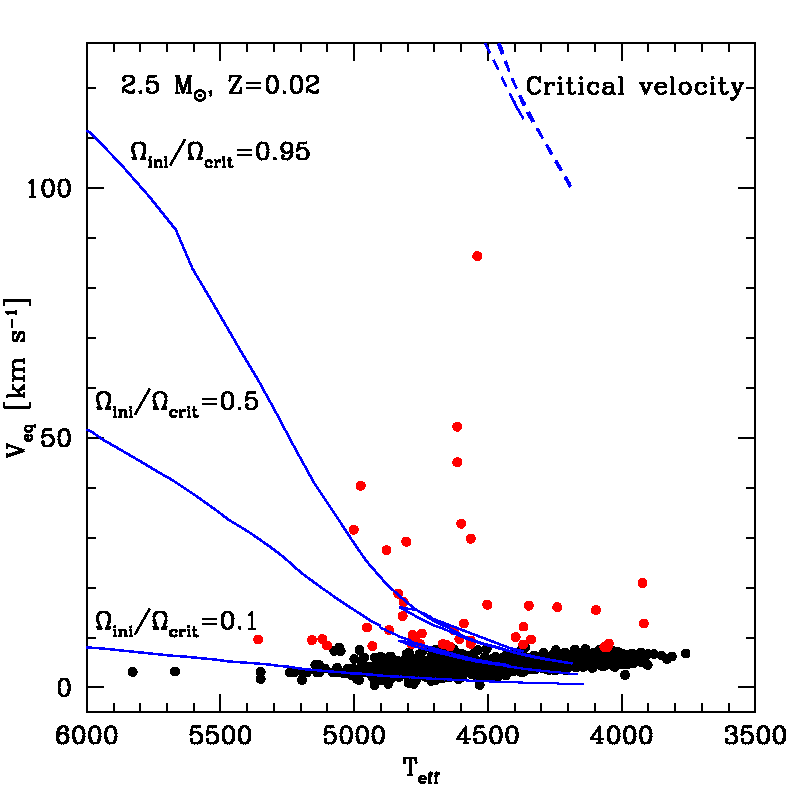}
\caption{Surface velocities $\upsilon\sin i$ as a function of the effective temperature for the sample of red giants observed by \citet{carlberg11}. The red dots correspond to stars with $\upsilon\sin i$ larger than 8 km s$^{-1}$. The lines indicate the evolutionary tracks for 2.5 M$_\odot$ stellar models for different initial rotations (from bottom to top, the time-averaged velocities during the MS phases are $\sim$20, 100 and 280 km s$^{-1}$). The upper dotted line shows the critical velocity.}
\label{fig:obs2011}
\end{figure}

The observed sample of \citet{carlberg11} (1287 stars) is shown in Fig.~\ref{fig:obs2011} superposed to the evolution of the surface velocities for different 2.5 M$_\odot$ models as computed in the present work. A significant number of stars are rotating much faster than what is predicted by these models. An extreme case is the star Tyc5854-011526-1 that has a $v \sin i$ of 84 km s$^{-1}$ which corresponds to 2/3 the critical velocity of a 2.5 M$_\odot$ at the considered effective temperature! 
In the following of this  section, we compare our theoretical predictions with the observations of red giants performed by \citet{Carlberg2012}. We choose this sample since it gives surface velocities and surface gravities for a significant number of red giants
(91 stars). Moreover, it provides also indications for some stars on the $^{12}$C/$^{13}$C ratio at the surface as well as for the lithium abundance allowing to check whether stars that are believed to have acquired their high surface rotation due to a planet engulfment
show also some special features in their surface composition.

We shall not perform a very detailed analysis since that analysis was already done in \citet{Carlberg2012}, but we want to address the following questions:
\begin{itemize}
\item Are there among the observed stars, cases that cannot be explained by single star models?
\item Can the observed surface velocities of such cases be reproduced by invoking a planet engulfment?
\item If yes, is it possible to deduce the initial conditions required to explain these systems by a planet engulfment?
\item Is there any chance to see some signatures of the planet engulfment in the surface composition?
\end{itemize}

Before investigating these points based on the present computations, we must see whether the metallicities and the mass range of our stellar models are adequate for a comparison with the sample of \citet{Carlberg2012}. For what concerns the metallicity, the correspondence between the metallicity of the observed sample and the metallicity of the models is marginal. The [Fe/H] of the sample studied by \citet{Carlberg2012} is on average lower than solar, while the present computations were performed with a heavy mass fraction of $Z=0.020$ which  is over solar ($Z_{\odot}=0.014$). However, 
whatever the metallicity, the engulfment would produce a strong acceleration of the surface. It may be that a given surface velocity would be reached with slightly different initial conditions but the main outcomes will not be changed. 

In the left panel of Fig.~\ref{comp1}, the observed sample is shown together with the tracks of the single (no engulfment) stellar models computed in the present work (case with $\Omega_{\rm ini}/\Omega_{\rm crit}=0.1$). 
In case we would have made our computations of stellar models with a lower initial metallicity, the tracks would have been slightly shifted to the left, {\it i.e.} to the blue. Keeping this in mind, we see that the range of masses followed here, would more or less go through the averaged observed positions. It may be that some stars have lower initial mass than 1.5 M$_\odot$ and some may have larger initial mass than 2.5 M$_\odot$, but on the whole the mass range of the models is relatively well representative of most of the masses of the sample. 

 \subsection{Stars with a clear signature of a past interaction}
 
 In many works, a red giant is considered as fast rotating and thus as being a possible candidate for having gone through a planet engulfment process if its
 $\upsilon\sin i$ is larger than 8 km s$^{-1}$. This criterion is however too schematic. To understand this point, let us look at the right panel of Fig.~\ref{comp1}. 
 We see that, actually, many of the {red}-filled points ($\upsilon\sin i >$8 km s$^{-1}$) are not very far from the line corresponding to the track for the 1.5 M$_\odot$ with $\Omega_{\rm ini}/\Omega_{\rm crit}=0.5$ allowing such stars to be possibly explained without invoking any interaction with a companion.
On the contrary, much lower surface velocities can be the indication of a planet engulfment. This can be seen looking at the magenta-dotted line which is in the lower-right corner on the right panel of Fig.~\ref{comp1}.
The tracks with no engulfment pass well below the slowly rotating points observed with surface gravities below about 1.6, while the track with engulfment of a one Jupiter mass planet initially at a distance of 0.5 au of a 1.5 M$_\odot$ star with $\Omega_{\rm ini}/\Omega_{\rm crit}=0.5$ would allow to go through at least part of these observed points.
Thus, the surface rotation above which some interaction may be needed depends strongly on the surface gravity and cannot be given by only one number.

How can we distinguish stars that would nearly certainly require some interaction as the cause for their high surface velocities?
The continuous (blue) line
shows the evolution of the surface velocity for a 2.5 M$_\odot$ that would begin its evolution at the critical velocity and evolves as a solid body rotating model. This evolution, as explained
in the previous section, represents an extremum in term of efficiency of the internal angular momentum transport and of the surface rotation.
Actually, this is a quite generous upper limit because we know that
red giants are not rotating as solid body. 

Turning now to stars that are found above this limit, we can be quite confident that, for these cases, an interaction
with a companion must have occurred. These stars are shown as circled blue magenta filled points and individually labeled by a capital letter. So, this answers
our first question above. We see that stars are indeed observed with surface velocities that cannot be explained by single star evolution\footnote{We may wonder, however, whether a more massive star (like a 3 M$_\odot$) could explain these points. At least this does not appear to be the case for point A which has, according to its position
in the theoretical HR diagram, an initial mass clearly below 2.5 M$_\odot$.}.

As a final remark, we note, as was already done in the introduction, that the bulk of stars with $\upsilon\sin i$ lower than 8 km s$^{-1}$ are in between the single star evolutionary tracks
with $\Omega_{\rm ini}/\Omega_{\rm crit}=0.1$ and $\Omega_{\rm ini}/\Omega_{\rm crit}=0.5$. This shows that single star models can well account for the surface rotation of the bulk of red giants.

\subsection{Can planet engulfment explain the cases likely resulting from an interaction?}

Can planet engulfment reproduce the surface velocities observed for 
 those red giants that have such large surface rotations that no
single star models can reproduce them (see the points labeled by high-case letters A to F in Figs.~\ref{comp2} and \ref{comp1})?
 The answer is yes, as can be seen
by comparing the locations of these stars in the right panel of Fig.~\ref{comp1} with the  black continuous tracks. Only a few cases are shown
for purpose of illustration, but there is no doubt that playing with the initial conditions (mass of the star, of the planet, its initial distance), it is indeed possible to produce a sufficient surface acceleration to reach these observed surface velocities and this for a sufficiently long duration to allow these high surface velocities to be observed.

\subsection{Can the initial conditions be deduced from observations?}

Can we determine the initial conditions needed to reproduce these systems? The solutions are unfortunately not unique. For instance, the positions of the points E and F could be reproduced by the engulfment of typically a 10 M$_{\rm J}$ planet initially at a distance of 0.5 au of a 2.5 M$_\odot$ with $\Omega_{\rm ini}/\Omega_{\rm crit}=0.5$, or by
the engulfment of a 10 M$_{\rm J}$ planet initially at a distance of 0.5 au from a 1.5 M$_\odot$ with $\Omega_{\rm ini}/\Omega_{\rm crit}=0.5$, and there are likely other solutions.
Of course, additional information can provide some further constraints. For instance, from the position in the HR diagram of the E and F stars (see left panel), the solution with the 2.5 M$_\odot$ would be favored with respect to the one involving a 1.5 M$_\odot$.
In that case the star would be likely in the clump and not along the red giant branch.

\subsection{The surface compositions after an engulfment}

\begin{figure*}
\includegraphics[width=0.49\textwidth, angle=0]{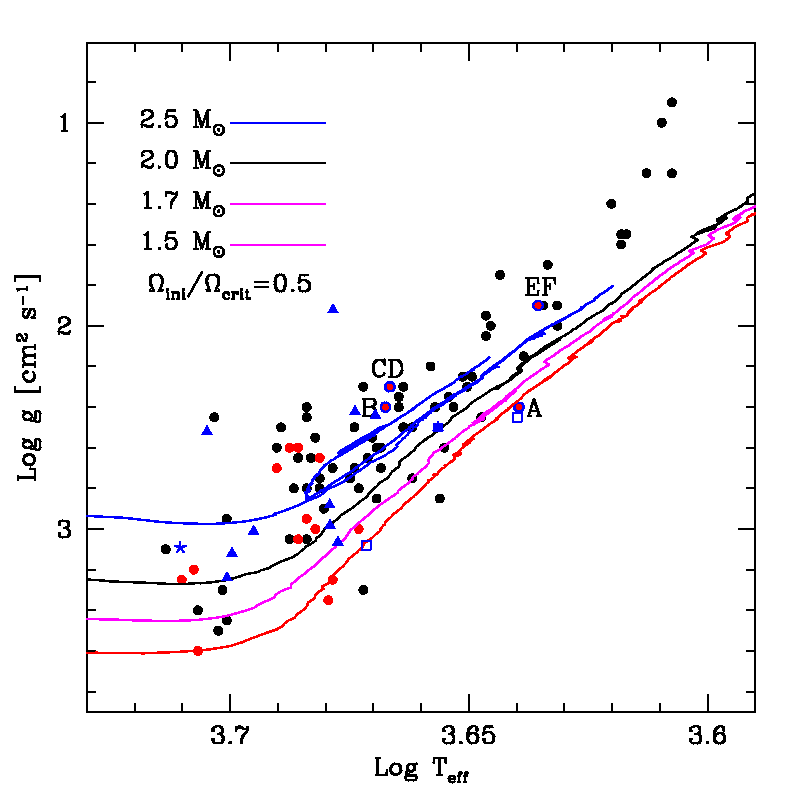}\includegraphics[width=0.49\textwidth, angle=0]{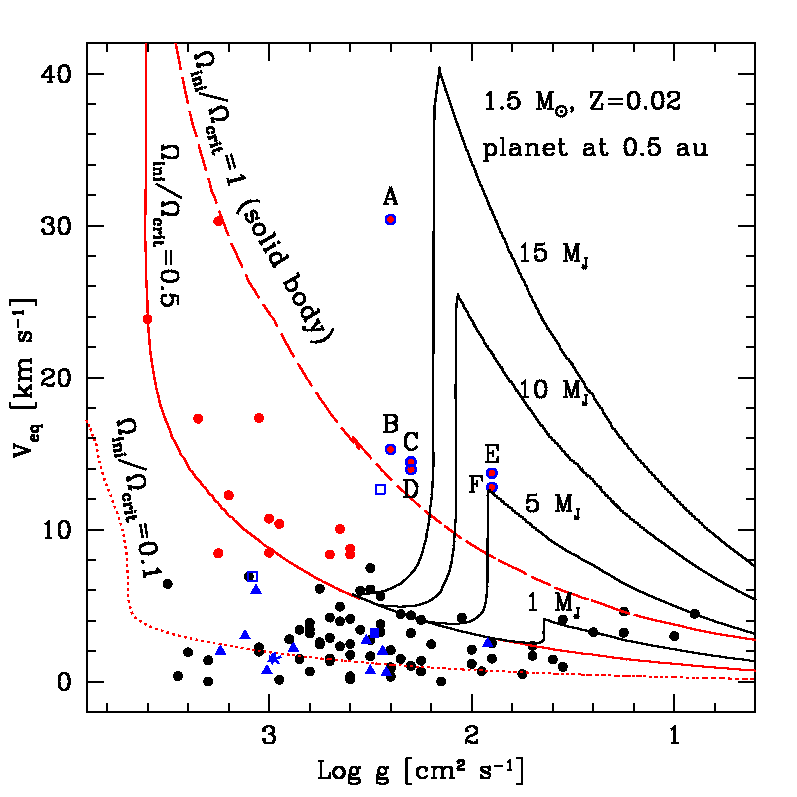}
\caption{{\it Left panel:} Surface gravities versus effective temperatures. The continuous lines correspond to evolutionary tracks for single star models (no engulfment) for (from bottom to top) 1.5 (red), 1.7 (magenta), 2.0 (black) and 2.5 M$_\odot$ (blue)
starting with an initial angular velocity equal to 50\% the critical one. The dots show the observations by \citet{Carlberg2012}, the black  circles are for stars with $\upsilon \sin i < 8$ km s$^{-1}$, the filled red and  circled blue magenta points are for stars 
with $\upsilon \sin i > 8$ km s$^{-1}$. The velocities of the  circled blue magenta points cannot be explained by any reasonable model for single stars. The high rotation of these stars results with great probability from an
interaction with an additional body. These points are labeled by letters (see text). The filled blue triangles are observations by \citet{Adamow2014}, the filled blues square is the Li-rich star BD+48 740 discussed by \citet{Adamow2012}, and the blue star is the Super Li-rich giant HD 107028 studied by \citet{Adamow2015}. The two empty squares correspond to fast rotators obtained by \citet{Tayar2015} for stars with masses equal to 1.43 and 2.07 M$_\odot$ along the red giant branch.
{\it Right panel:} Surface equatorial rotation versus surface gravities for 1.5 M$_\odot$ models. The two red tracks correspond to no planet engulfment cases starting with different initial rotations. 
The black curves show the evolution for stars with planet engulfment. The dots are the same observations as those shown in the left panel.
}
\label{comp1}
\end{figure*} 

\begin{figure*}
\includegraphics[width=0.49\textwidth, angle=0]{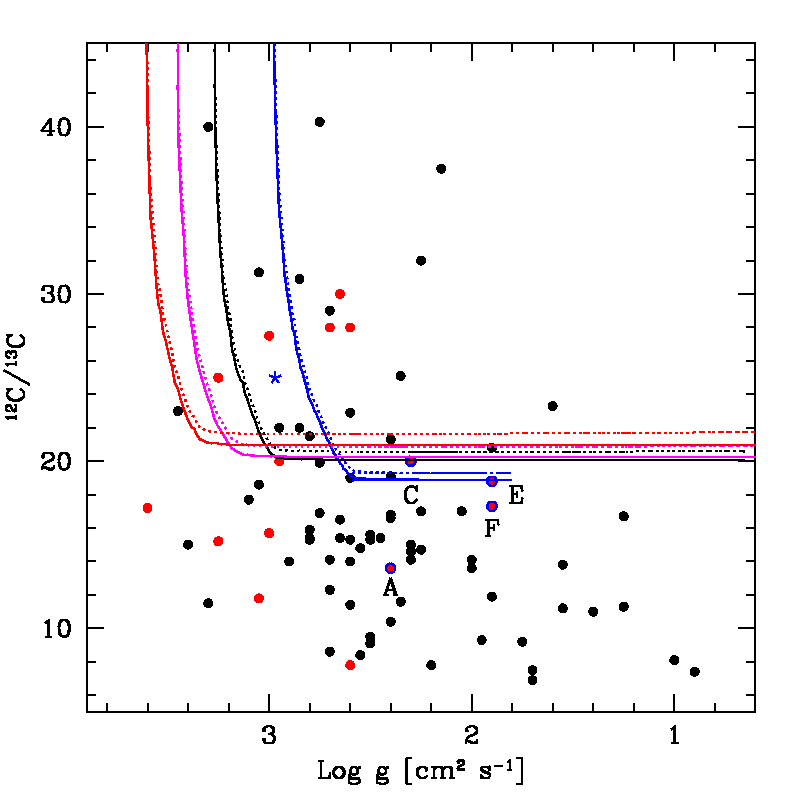}\includegraphics[width=0.49\textwidth, angle=0]{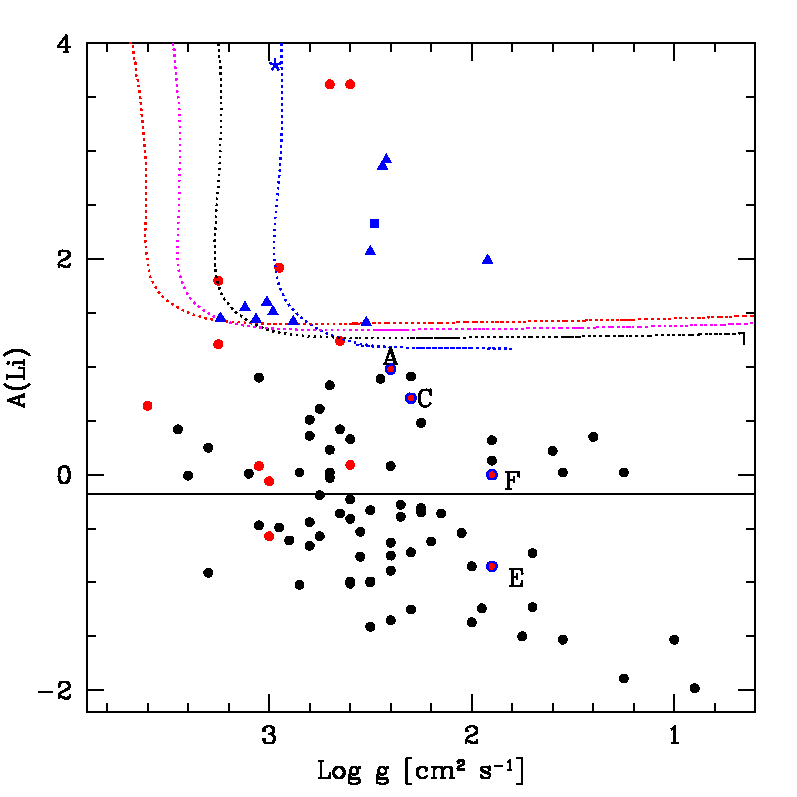}
\caption{Comparison between surface abundances and theoretical predictions. The points are observations (see caption of Fig.~\ref{comp1} for the meaning of the different symbols).
{\it Left panel:} Surface isotopic ratios $^{12}$C/$^{13}$C versus surface gravities. 
The continuous lines are the same evolutionary tracks as those shown in the left panel of Fig.~\ref{comp1}. The dotted lines
show the new isotopic ratios obtained when account is made for the dilution in the convective envelope of the planet material (see text for more details) starting from the models
with $\Omega_{\rm ini}/\Omega_{\rm crit}=0.5$.
{\it Right panel:} Surface lithium abundance versus surface gravities. The continuous (black) horizontal line shows the averaged lithium abundance observed at the surface
of stars with $\upsilon\sin i$ smaller than 8 km s$^{-1}$. The dotted lines show the results of dilution of the planet material into the convective envelope.
}
\label{comp2}
\end{figure*}

The left panel of Fig.~\ref{comp2} shows the isotopic ratios at the surface of the observed stars as well as the predictions of the theoretical models. The continuous lines show
the predictions of the present stellar models without any planet engulfment. 
The predictions depend on the initial rotation. Larger initial rotations lead to lower $^{12}$C/$^{13}$C ratios. Models with an initial rotation equal to 50\% the critical one 
go roughly through the middle of the observed points\footnote{The present models
do not account for thermohaline mixing for instance, which has been proposed by \citet{CZ2007} as causing an additional decrease of the
$^{12}$C/$^{13}$C ratio and of the lithium along the red giant branch \citep[see also][]{CL2010}.}.

We can deduce the following points looking at the left panel of Fig.~\ref{comp2}:
\begin{itemize}
\item  The planet engulfment process decreases 
the shear between the core and the convective envelope and thus weakens the shear mixing in that region.
However this effect is negligible because, even in models without engulfment, which have a stronger shear in that region, we see
no effect. Indeed, such an effect would produce a continuous decrease of the
$^{12}$C/$^{13}$C ratio once the convective envelope has reached its deepest point. This is obviously not the case as shown in 
the left panel of Fig.~\ref{comp2}.
\item Another process can change the isotopic ratio: the dilution of the planet material into the convective envelope. To estimate the importance of this process,
we have followed the same line of reasoning as presented in \citet{Carlberg2012}. We applied their equation (3), which gives the new isotopic ratio, $(^{12}C/^{13}C)_{\rm new}$ accounting for the dilution of the planetary material into the convective envelope:
$$
(^{12}C/^{13}C)_{\rm new}
=
{
10^{A(C)_p} {r_p q_e \over 1+r_p}
+
10^{A(C)_*} {r_* \over 1+r_*}
\over
10^{A(C)_p} {q_e\over 1+r_p}
+
10^{A(C)_*} {1 \over 1+r_*}
}
$$
where $A(C)_p$ is Log$(N(C)/N(H))+12$ in the planet and $A(C)_*$ is the same quantity for the star
($N(C)$ is the number of carbon per unit volume accounting for the two isotopes $^{12}$C and $^{13}$C).
The ratio $r$ is equal to $N(^{12}C)/N(^{13}C)$, and $q_e$ is the ratio of the mass of the planet
to the mass of the convective envelope. For the quantities concerning the star ($*$) we took the values
directly from our stellar models. For the planet, we considered the same values as in \citet{Carlberg2012}
namely $A(C)_p=8.87$ dex \citep[about three times the solar carbon abundance as determined by][]{Wong2004}
and $r_p=89$ which is a standard value for the solar system \citep{Lodders1998}. Using the above formula, we obtain after engulfment the level of $^{12}C/^{13}C$ shown by the dashed segments
in Fig.~\ref{comp2}. We see that since the planetary material has a higher $^{12}C/^{13}C$ ratio than the stellar envelope,
this effect makes the surface isotopic ratio a bit larger. The effect is however small and thus the measure of this isotopic ratio
appears as a poor indicator of a planet engulfment, especially that changes of the initial mass, rotation and likely other possible mixing processes have stronger effects
than the engulfment.
\item Looking only at the observed points, we see that
the fast rotators and those stars whose surface velocities cannot be explained by single star evolution do not present different carbon isotopic ratios than the rest of the sample, confirming
that this isotopic ratio does not appear to be very sensitive to the engulfment.
\end{itemize}

In the right panel of Fig.~\ref{comp2}, observed lithium abundances are shown. It seems as indicated by \citet{Carlberg2012} that the fast rotators present on average higher lithium abundances.
However, if we consider the Li-rich giants observed by \citet{Adamow2012, Adamow2014, Adamow2015}, no one of these stars has a $\upsilon\sin i > 8$ km s$^{-1}$ (see the right panel of Fig.~\ref{comp1})
somewhat blurring the trend. 

Similar estimates as the one done above for the carbon isotopic ratio can be performed for the lithium abundance. Using equation (2) of \citet{Carlberg2012}
$$
A(Li)_{\rm new}={\rm Log}(q_e 10^{A(Li)_p}+10^{A(Li)_*})-{\rm Log(1+q_e)}
$$
we estimate the new surface lithium abundance, $A(Li)_{\rm new}$, after accretion of the planetary material. In our stellar models,
the abundance of lithium is not followed explicitly, thus we did as in \citet{Carlberg2012}, considering for 
 $A(Li)_{*}$ a value equal to -0.18 dex corresponding to the averaged abundance of lithium observed at the surface of slow rotators (see the horizontal
 continuous line in the right panel of Fig.~\ref{comp2}). For $A(Li)_{\rm p}$, we took a value equal to 3.3 \citep{Lodders1998}, while for $q_e$, we used
 the stellar models quantities. The results are shown by the dotted tracks in the right panel of Fig.~\ref{comp2}.  The lines represent the surface lithium abundances
 that would be observed in case the engulfment would occur at various surface gravities. When
 the outer convective zone begins to appear, the convective envelope is very small and the Li abundance that results from a planet dilution is very high. 
 This makes the vertical part of the track. Then when the envelope increases in mass, the evolution goes to lower Li abundances and to lower gravities.
 
 From the right panel of Fig.~\ref{comp2},  we see that
 indeed the dilution of the planetary material into the convective envelope can have a strong effect on the surface Li abundance. 
 The conditions at the base of the convective of our stellar models are not favorable to a rapid destruction of that element by proton captures, thus
 it will not be burned unless some mixing occurs below the convective envelope \citep[see also][]{Aguilera2016}.
 
 The strong effect on lithium abundance found here after the planet engulfment confirms what was suggested by many authors in the past \citep[see {\it e.g.}][]{alexander67, siess99I, Carlberg2012,Adamow12}, namely that
 the planet engulfment process can indeed produce lithium-rich red giants. However, the  lithium-signature of planet engulfment remains difficult to disentangle from other
 processes. Lithium is a fragile element destroyed at about 2.6 million degrees, so any surface enrichment might in some stellar models disappear rapidly. Lithium can
 also be produced in some red giant models blurring completely the picture for what concerns the origin of a high lithium abundance.
 
 In that respect it is interesting to see what are the lithium surface abundances of those stars whose high surface velocities cannot be explained by single star models
 (see the circled blue magenta filled points in Fig.~\ref{comp2}). 
  Their positions are not very peculiar in the sense that slowly rotating stars 
 are also found with similar levels of Li abundances. 
  However we see that  all these stars are above the averaged level for the (apparently) slowly rotating stars, leaving open the possibility that they could have received some lithium from a planet.
  
  On the other hand, as noted already above, none of the Li-rich giants observed by \citet{Adamow2012, Adamow2014, Adamow2015} are apparently fast rotators. These kind of stars are therefore
  difficult to be explained by a planet engulfment process unless some mechanism, not accounted for in the present models, as a strong wind magnetic braking for instance has occurred. In that
  case those red giants would have been once fast rotators, but the wind magnetic braking was efficient enough for the star to have lost memory of its  post-engulfment fast rotating stage.

\section{Perspectives and Conclusions}\label{sec:7}

The main conclusions of the present paper can be summarized in the following way:
\begin{itemize}
\item As was already shown in \citet{paperI}, there are observed red giant stars whose high surface velocities cannot be explained by single star evolution.
The upper velocities that can be reached by single star models of the considered mass are indicated as a function of the surface gravity in Fig.~\ref{fig:vlog} by the
long-dashed black line labeled V$_{\rm max}$. The use of such gravity dependent limits to isolate red giants having likely engulfed a planet  or at least interact with it is probably more
realistic than the commonly used limit of 8 km s$^{-1}$.
\item The present models show that tidal interactions followed by planet engulfment can reproduce the high 
surface velocities of these stars during sufficiently long periods for allowing these high rotations to be observable.
Surface velocities beyond the upper limit
allowed by single star evolution can be reached already just by tidal interaction.
\item We obtain also that the high rotation obtained after the engulfment is maintained beyond the end of the red giant branch  for the 2.5 M$_\odot$.
It would be extremely interesting to check whether such a conclusion would hold for our lower initial mass models that go through a He-flash episode.
\item Conditions also exist for producing engulfments with much less spectacular impacts. Let us remind that
the engulfment of a one M$_{\rm J}$ planet by a 1.5 M$_\odot$ star would produce surface velocities of only
a few km per seconds at gravities below about 1.6. This is quite small and these stars would hardly be recognized as having engulfed a planet.
\item The above conclusions make the link between observed fast-rotating red giants and red giants having engulfed a planet much less clear, in the sense
that not all fast rotating red giants need an interaction to have reached their surface velocities (this may reflect simply a high initial rotation rate of the
progenitor) and not all slowly rotating red giants may be explained by single star evolution (typically in the upper part of the red giant branch).
\item We discussed in a simple way the consequences of an engulfment on the surface composition of red giants
and confirmed the results obtained by \citet{Carlberg2012} that the effect on the carbon isotopic ratios are very small and the impact on lithium 
might be quite large, although remaining difficult to interpret (see below). 
\item We showed that the chemical signatures of an engulfment are still quite ambiguous because
planet engulfment either produce too small signal as in the case of the carbon isotopic ratios, or produce signals
that cannot be attributed in a non ambiguous way to a planet engulfment.
{\it This makes the surface velocity of a red giant the most stringent observable feature indicating a past star-planet interaction.}
\item The present results also show that the evolution after an engulfment is not very different
from that obtained without any engulfment. The main difference is in the surface rotation (and of course the rotation as a whole
of the external convective zone). A planet engulfment would somewhat lower the contrast between the angular velocity of the core and that of the envelope
but not at a level that could be useful to reproduce the small contrast as obtained by asteroseismology for a few red giants \citep[e.g.][]{beck12,Mosser2012,deh12,deh14}.
\end{itemize}
There are many other points that should be addressed in further works. In particular, we shall investigate whether the acceleration of the surface
due to a planet engulfment may trigger a magnetic field.

\begin{acknowledgements} 
This research has made use of the Exoplanet Orbit Database
and the Exoplanet Data Explorer at exoplanets.org. The project has been supported by Swiss National Science Foundation grants 200021-138016,  200020-160119 and 200020-15710.
\end{acknowledgements}

\begin{appendix} \label{sec:appe}
\section{Detailed properties of the stellar models before and after engulfment}

This appendix contains three tables giving information on the characteristics of the star before and after an engulfment. Table~\ref{table:stru0.1} indicates the properties of the stellar models
just before the engulfment. 
The first column is the mass of the planet. Columns 2 to 5 give the mass of the star, its radius, surface angular velocity, mass loss rate and mass of the convective envelope at this time.

\begin{figure}
\includegraphics[width=.49\textwidth, angle=0]{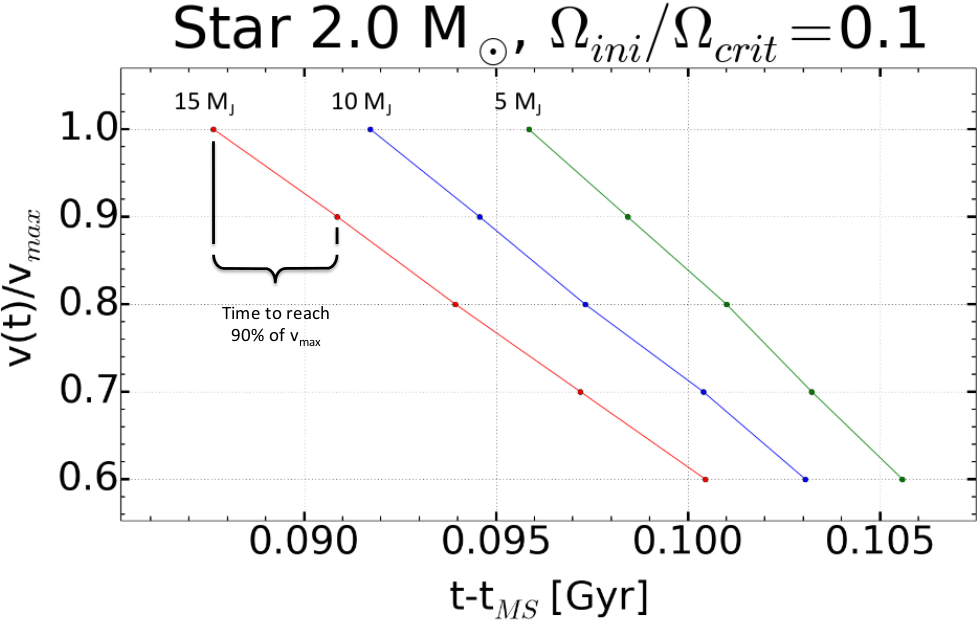}
\caption{Evolution of the surface velocity (normalized to the velocity reached just after the engulfment, $v_{\rm max}$) as a function of time for a 2 M$_\odot$ stellar model with an initial rotation equal to 10\% the critical velocity. Various masses for the planets are considered. All planets started at a distance equal to 0.5 au from their parent star.}
\label{fig:decrease}
\end{figure}

Tables \ref{table:Omega0.1} and \ref{table:Omega0.5} present the changes occurring due to the tides/engulfment processes. 
The first column is the mass of the planet. Columns 2 gives the equatorial surface velocity before the engulfment. Note that this velocity
is already significantly higher than the one of the corresponding {\it single} star at that stage. Actually, the velocity indicated in that column
would be the velocity acquired by the tidal forces only.
The duration of the phases 
during which, starting from the velocity indicated in column 2, 
the surface velocity would be higher than 8 km s$^{-1}$, $\Delta t_t(FR)$, is given in column 3.
To compute this duration we used the results shown in Fig.~\ref{fig:decrease}.
This figure shows the evolution as a function of time of the 
surface velocity just after an engulfment. We see that roughly the surface velocity varies as
$v(\Delta t) \sim v_{\rm max}(1-0.04 \Delta t)$, 
where  $\Delta t$ (in Myr) is $t-t_{\rm MS}$, with $t$ the age of the star, $t_{\rm MS}$ the age of the star at the end of the MS phase, and $v_{\rm max}$ the surface
velocity reached just after the engulfment.
This relation does not much depend on the mass of the planet, or equivalently on the value of $v_{\rm max}$.
It means that after 2.5 Myr, whatever the initial velocity,
the surface velocity will have decreased by 10\%
with respect to its initial value\footnote{Actually, this timescale changes a bit with the mass of the planet as can be seen in Fig.~\ref{fig:decrease}, being slightly shorter for the 5 than for the 15 M$_{\rm J}$ planet. On the other hand the differences between the various planet masses are not very large.}. We can use such a plot to estimate the time during which a star will maintain a certain velocity after the engulfment evolving up along the red giant branch. This is how we have estimated the
time, $\Delta t_{t}(FR)$, indicated in column 3 of Tables \ref{table:Omega0.1} and \ref{table:Omega0.5} showing the duration of the period when the star
has a surface velocity higher than 8 km s$^{-1}$ when only the acceleration due to tides is accounted for.

In Tables \ref{table:Omega0.1} and \ref{table:Omega0.5}, columns 4 to 6 indicate
the surface velocity right after the engulfment, the duration of the phases 
during which the surface velocity, after an engulfment, is higher than 8 km s$^{-1}$ ($\Delta t_{\rm t+e}(FR)$), and finally the fraction
of the red giant branch phase spent with a surface velocity superior to 8 km s$^{-1}$. Tables \ref{table:Omega0.1}
show the results obtained with stellar models having an initial angular velocity equal to 10\% the critical velocity, Tables \ref{table:Omega0.5}
the results for models having an initial angular velocity equal to 50\% the critical velocity.

\begin{table*}
\scriptsize{
\caption{Characteristics of the star at the time of engulfment, see also Paper I for additional pieces of information. A small $-$ indicates that no engulfment occurs during the red giant phase for the
considered initial conditions.} 
\resizebox{18cm}{!} {
\begin{tabular}{c|ccccc||ccccc} 
\hline\hline 
$M_{\rm pl}$ & $M_{\star}(Age_{\rm eng})$ & $R_{\star}(Age_{\rm eng})$ & $\Omega_{\rm surf}(Age_{\rm eng})$ & $\dot{M}_{\star}(Age_{\rm eng})$ & $M_{\rm env}(Age_{\rm eng})$ & $M_{\star}(Age_{\rm eng})$ &  $R_{\star}(Age_{\rm eng})$ & $\Omega_{\rm surf}(Age_{\rm eng})$  & $\dot{M}_{\star}(Age_{\rm eng})$ & $M_{\rm env}(Age_{\rm eng})$\\ 
 $[M_{\rm J}]$ & $[M_{\odot}]$ & $[R_{\odot}]$ & [10$^{-7}$ s$^{-1}$] & $[10^{-9} M_{\odot}yr^{-1}]$  & [$\%$$M_{\star}$] &  $[M_{\odot}]$ & $[R_{\odot}]$ & [10$^{-7}$ s$^{-1}$] & $[10^{-9} M_{\odot}yr^{-1}]$    & [$\%$$M_{\star}$]  \\
\hline 
\multicolumn{1}{c|}{ }&\multicolumn{10}{c}{$\Omega_{\rm ini}$/ $\Omega_{\rm crit}$ = 0.1, Z = 0.02}\\
\cline{2-11}
 \multicolumn{1}{c|}{ }&\multicolumn{5}{c||}{}& \multicolumn{5}{c}{}\\
\multicolumn{1}{c|}{ }&\multicolumn{5}{c||}{$M_{\star,ini}=1.5\ M_{\odot}$, $R_{\star,ini}=1.506\ R_{\odot}$, $v_{\rm surf,ini}=25.6\ km s^{-1}$}&\multicolumn{5}{c}{$M_{\star,ini}=1.7\ M_{\odot}$, $R_{\star,ini}=1.535\ R_{\odot}$, $v_{surf,ini}=25.1\ km s^{-1}$}\\
 \multicolumn{1}{c|}{ }&\multicolumn{5}{c||}{}& \multicolumn{5}{c}{}\\
\hline 
\multicolumn{1}{c|}{ }&\multicolumn{10}{c}{a$_{\rm ini} = 0.5$ [AU]}\\
\hline 
1	&	1.485274	&	29.3838	&	1.01	&	-0.77	&	67.62	&	1.687587	&	30.5011	&	0.84	&	-0.79	&	67.09	\\
5	&	1.490112	&	20.8069	&	9.55	&	-0.32	&	69.45	&	1.691796	&	22.2625	&	5.58	&	-0.35	&	68.87	\\
10	&	1.492073	&	17.4992	&	23.50	&	-0.21	&	70.46	&	1.693976	&	17.6225	&	15.97	&	-0.19	&	69.85	\\
15	&	1.493253	&	15.4550	&	38.69	&	-0.16	&	71.03	&	1.694331	&	17.1070	&	29.72	&	-0.18	&	70.01	\\
\hline 
\multicolumn{1}{c|}{ }&\multicolumn{10}{c}{a$_{\rm ini} = 1.0$ [AU]}\\
\hline 
1	&	1.458215	&	70.2482	&	0.26	&	-6.75	&	61.94	&	1.663381	&	70.2862	&	0.13	&	-6.38	&	61.41	\\
5	&	1.472279	&	50.4068	&	2.10	&	-3.02	&	64.37	&	1.675412	&	52.1931	&	1.73	&	-3.08	&	63.64	\\
10	&	1.477916	&	42.1161	&	5.87	&	-2.00	&	65.56	&	1.680259	&	44.1649	&	4.67	&	-2.07	&	64.88	\\
15	&	1.481351	&	37.0043	&	10.69	&	-1.48	&	66.52	&	1.683111	&	38.9341	&	9.03	&	-1.55	&	65.66	\\
\hline 
\multicolumn{1}{c|}{ }&\multicolumn{10}{c}{a$_{\rm ini} = 1.5$ [AU]}\\
\hline 
1	&	1.422241	&	109.5148	&	0.15	&	-20.61	&	57.76	&	1.626031	&	115.7902	&	0.12	&	-21.88	&	56.75	\\
5	&	1.448623	&	81.8077	&	1.06	&	-9.98	&	60.60	&	1.652720	&	85.0667	&	0.99	&	-10.21	&	59.84	\\
10	&	1.459680	&	68.0247	&	2.00	&	-6.38	&	62.19	&	1.663087	&	71.0516	&	1.14	&	-6.56	&	61.42	\\
15	&	1.465830	&	62.7179	&	5.29	&	-5.69	&	63.21	&	1.668403	&	64.7079	&	4.11	&	-5.42	&	62.29	\\
\hline 
 \multicolumn{1}{c|}{ }&\multicolumn{5}{c||}{}& \multicolumn{5}{c}{}\\
\multicolumn{1}{c|}{ }&\multicolumn{5}{c||}{$M_{\star,ini}=2.0\ M_{\odot}$, $R_{\star,ini}=1.620\ R_{\odot}$, $v_{\rm surf,ini}=26.5\ km s^{-1}$}&\multicolumn{5}{c}{$M_{\star,ini}=2.5\ M_{\odot}$, $R_{\star,ini}=1.785\ R_{\odot}$, $v_{\rm surf,ini}=28.2\ km s^{-1}$}\\
 \multicolumn{1}{c|}{ }&\multicolumn{5}{c||}{}& \multicolumn{5}{c}{}\\
\hline 
\multicolumn{1}{c|}{ }&\multicolumn{10}{c}{a$_{\rm ini} = 0.5$ [AU]}\\
\hline 
1	&	1.989876	&	34.5306	&	0.52	&	-0.97	&	65.38	&	-	&	-	&	-	&	-	&	-	\\
5	&	1.993967	&	24.2140	&	4.58	&	-0.39	&	66.90	&	2.498326	&	31.0165	&	2.32	&	-0.65	&	63.77	\\
10	&	1.995228	&	20.4300	&	11.68	&	-0.26	&	67.48	&	2.498676	&	28.5510	&	3.78	&	-0.52	&	63.77	\\
15	&	1.995974	&	18.1078	&	17.46	&	-0.19	&	67.86	&	2.498689	&	28.4553	&	6.89	&	-0.52	&	63.77	\\
\hline 
\multicolumn{1}{c|}{ }&\multicolumn{10}{c}{a$_{\rm ini} = 1.0$ [AU]}\\
\hline 
1	&	1.969336	&	73.7458	&	0.19	&	-6.52	&	60.02	&	-	&	-	&	-	&	-	&	-	\\
5	&	1.979736	&	55.3260	&	1.41	&	-3.26	&	62.39	&	-	&	-	&	-	&	-	&	-	\\
10	&	1.983725	&	47.8862	&	3.63	&	-2.30	&	63.30	&	-	&	-	&	-	&	-	&	-	\\
15	&	1.986025	&	43.0656	&	5.95	&	-1.76	&	64.05	&	-	&	-	&	-	&	-	&	-	\\
\hline 
\multicolumn{1}{c|}{ }&\multicolumn{10}{c}{a$_{\rm ini} = 1.5$ [AU]}\\
\hline 
1	&	1.936279	&	119.2380	&	0.06	&	-21.53	&	55.27	&	-	&	-	&	-	&	-	&	-	\\
5	&	1.959307	&	89.1782	&	0.68	&	-10.47	&	58.46	&	-	&	-	&	-	&	-	&	-	\\
10	&	1.967829	&	76.6386	&	1.72	&	-7.29	&	59.86	&	-	&	-	&	-	&	-	&	-	\\
15	&	1.972431	&	68.9721	&	3.21	&	-5.66	&	60.75	&	-	&	-	&	-	&	-	&	-	\\
\hline 
\multicolumn{1}{c|}{ }&\multicolumn{10}{c}{$\Omega_{\rm ini}$/ $\Omega_{\rm crit}$ = 0.5, Z = 0.02}\\
\cline{2-11}
 \multicolumn{1}{c|}{ }&\multicolumn{5}{c||}{}& \multicolumn{5}{c}{}\\
\multicolumn{1}{c|}{ }&\multicolumn{5}{c||}{$M_{\star,ini}=1.5\ M_{\odot}$, $R_{\star,ini}=1.473\ R_{\odot}$, $v_{\rm surf,ini}=142\ km s^{-1}$}&\multicolumn{5}{c}{$M_{\star,ini}=1.7\ M_{\odot}$, $R_{\star,ini}=1.580\ R_{\odot}$, $v_{\rm surf,ini}=130\ km s^{-1}$}\\
 \multicolumn{1}{c|}{ }&\multicolumn{5}{c||}{}& \multicolumn{5}{c}{}\\
\hline 
\multicolumn{1}{c|}{ }&\multicolumn{10}{c}{a$_{\rm ini} = 0.5$ [AU]}\\
\hline 
1	&	1.485476	&	30.4342	&	1.93	&	-0.86	&	67.47	&	1.688074	&	31.4697	&	1.56	&	-0.87	&	66.58	\\
5	&	1.490038	&	22.1058	&	8.21	&	-0.38	&	69.16	&	1.691756	&	23.5966	&	6.98	&	-0.42	&	68.03	\\
10	&	1.491929	&	18.6698	&	19.48	&	-0.25	&	70.01	&	1.693187	&	20.1127	&	14.86	&	-0.28	&	68.68	\\
15	&	1.492999	&	16.3263	&	35.18	&	-0.19	&	70.59	&	1.694140	&	17.6355	&	28.11	&	-0.21	&	69.17	\\		\hline 
\multicolumn{1}{c|}{ }&\multicolumn{10}{c}{a$_{\rm ini} = 1.0$ [AU]}\\
\hline 
1	&	1.459109	&	70.9117	&	0.44	&	-7.05	&	61.77	&	1.664231	&	72.3823	&	0.40	&	-7.01	&	61.06	\\
5	&	1.472818	&	51.5472	&	2.34	&	-3.24	&	64.21	&	1.676070	&	54.2353	&	1.92	&	-3.45	&	63.29	\\
10	&	1.482561	&	36.0369	&	8.27	&	-1.36	&	66.49	&	1.680738	&	45.5303	&	4.98	&	-2.30	&	64.37	\\
15	&	1.481257	&	37.5313	&	10.94	&	-1.60	&	66.23	&	1.683347	&	43.9700	&	12.58	&	-2.42	&	65.15	\\
\hline 
\multicolumn{1}{c|}{ }&\multicolumn{10}{c}{a$_{\rm ini} = 1.5$ [AU]}\\
\hline 
1	&	1.418420	&	116.2842	&	0.20	&	-24.21	&	57.16	&	1.627502	&	118.5560	&	0.18	&	-23.82	&	56.38	\\
5	&	1.449313	&	83.4278	&	0.63	&	-10.54	&	60.43	&	1.654111	&	87.8077	&	0.96	&	-11.17	&	59.47	\\
10	&	1.460203	&	69.9438	&	1.32	&	-6.84	&	62.03	&	1.663785	&	73.3716	&	1.13	&	-7.28	&	60.90	\\
15	&	1.466107	&	61.3007	&	5.41	&	-5.40	&	62.91	&	1.669147	&	65.6610	&	1.88	&	-5.55	&	61.94	\\
\hline 
 \multicolumn{1}{c|}{ }&\multicolumn{5}{c||}{}& \multicolumn{5}{c}{}\\
\multicolumn{1}{c|}{ }&\multicolumn{5}{c||}{$M_{\star,ini}=2.0\ M_{\odot}$, $R_{\star,ini}=1.667\ R_{\odot}$, $v_{\rm surf,ini}=138\ km s^{-1}$}&\multicolumn{5}{c}{$M_{\star,ini}=2.5\ M_{\odot}$, $R_{\star,ini}=1.833\ R_{\odot}$, $v_{\rm surf,ini}=147\ km s^{-1}$}\\
 \multicolumn{1}{c|}{ }&\multicolumn{5}{c||}{}& \multicolumn{5}{c}{}\\
\hline 
\multicolumn{1}{c|}{ }&\multicolumn{10}{c}{a$_{\rm ini} = 0.5$ [AU]}\\
\hline 
1	&	1.991163	&	33.5930	&	1.41	&	-0.93	&	64.96	&	-	&	-	&	-	&	-	&	-	\\
5	&	1.994217	&	25.0269	&	5.53	&	-0.44	&	66.10	&	2.498400	&	32.5785	&	3.10	&	-0.75	&	62.02	\\
10	&	1.996013	&	18.3343	&	16.65	&	-0.21	&	66.67	&	2.498741	&	29.5638	&	6.37	&	-0.59	&	62.02	\\
15	&	1.997273	&	21.2124	&	17.83	&	-0.29	&	68.44	&	2.498893	&	28.3750	&	9.41	&	-0.54	&	61.77	\\
\hline 
\multicolumn{1}{c|}{ }&\multicolumn{10}{c}{a$_{\rm ini} = 1.0$ [AU]}\\
\hline 
1	&	1.970755	&	75.3894	&	0.37	&	-7.06	&	59.60	&	-	&	-	&	-	&	-	&	-	\\
5	&	1.980597	&	57.0608	&	1.93	&	-3.61	&	61.77	&	-	&	-	&	-	&	-	&	-	\\
10	&	1.984246	&	49.5622	&	3.61	&	-2.56	&	62.70	&	-	&	-	&	-	&	-	&	-	\\
15	&	1.986321	&	46.0991	&	6.57	&	-2.15	&	63.25	&	-	&	-	&	-	&	-	&	-	\\
\hline 
\multicolumn{1}{c|}{ }&\multicolumn{10}{c}{a$_{\rm ini} = 1.5$ [AU]}\\
\hline 
1	&	-	&	-	&	-	&	-	&	-	&	-	&	-	&	-	&	-	&	-	\\
5	&	1.960723	&	92.4787	&	0.78	&	-11.56	&	57.84	&	-	&	-	&	-	&	-	&	-	\\
10	&	1.968937	&	78.6476	&	1.86	&	-7.91	&	59.24	&	-	&	-	&	-	&	-	&	-	\\
15	&	1.973119	&	71.6140	&	2.24	&	-6.30	&	60.14	&	-	&	-	&	-	&	-	&	-	\\
\hline 
\end{tabular}
\label{table:stru0.1}
}}
\end{table*}	

\begin{table*}
\scriptsize{

\caption{Effect of planet engulfment on the surface rotation velocities of red giants. The cases for different initial mass stars beginning their evolution
with a rotation equal to $\Omega_{\rm ini}$/ $\Omega_{\rm crit}$ = 0.1 are considered.}

\begin{tabular}{c|ccccc||ccccc} 

\hline\hline 
$M_{\rm pl}$ &   v$_{\rm surf}({\rm Be})$ & $\Delta t_{\rm t}({\rm FR})$  & v$_{\rm surf}({\rm Af})$  & $\Delta t_{\rm t+e}({\rm FR})$ & $\Delta t_{\rm t+e}({\rm FR})/t({\rm RG})$ &    v$_{\rm surf}({\rm Be})$ & $\Delta t_{\rm t}({\rm FR})$ & v$_{\rm surf}({\rm Af})$  & $\Delta t_{\rm t+e}({\rm FR})$ & $\Delta t_{\rm t+e}({\rm FR})/t({\rm RG})$\\ 
 $[M_{J}]$ & [km s$^{-1}$]  & [Myr]  & [km s$^{-1}$] & [Myr]  & [$\%$$t_{rgb}$] &  [km s$^{-1}$]& [Myr]   & [km s$^{-1}$] & [Myr]  & [$\%$$t_{rgb}$] \\
\hline 
\multicolumn{1}{c|}{ }&\multicolumn{10}{c}{$\Omega_{ini}$/ $\Omega_{crit}$ = 0.1, Z = 0.02}\\
\hline 
 \multicolumn{1}{c|}{ }&\multicolumn{5}{c||}{}& \multicolumn{5}{c}{}\\
\multicolumn{1}{c|}{ }&\multicolumn{5}{c||}{1.5 M$_{\odot}$, $\mathcal{L}_{0,\star} = 0.41$ [10$^{50}$ g cm$^2$ s$^{-1}$], $t_{rgb} \simeq $ 0.37 [Gyr]}
& \multicolumn{5}{c}{1.7 M$_{\odot}$, $\mathcal{L}_{0,\star} = 0.46$  [10$^{50}$ g cm$^2$ s$^{-1}$], $t_{rgb} \simeq $ 0.18 [Gyr]}\\
 \multicolumn{1}{c|}{ }&\multicolumn{5}{c||}{}& \multicolumn{5}{c}{}\\
\hline 
\multicolumn{1}{c|}{ }&\multicolumn{10}{c}{a$_{ini} = 0.5$ [AU]}\\
\hline 
1	&	1.44	&	-	&	2.06	&	0.0	&	0.00	&	1.35	&	-	&	1.8	&	0.0	&	0.00	\\
5	&	7.26	&	-	&	13.80	&	15.3	&	4.29	&	6.26	&	-	&	8.64	&	2.8	&	1.55	\\
10	&	17.00	&	26.24	&	28.90	&	29.6	&	8.30	&	14.60	&	19.65	&	19.60	&	23.9	&	13.13\\	
15	&	28.20	&	32.36	&	42.30	&	39.7	&	11.15	&	23.30	&	26.28	&	35.70	&	31.2	&	17.20\\	
\hline 
\multicolumn{1}{c|}{ }&\multicolumn{10}{c}{a$_{ini} = 1$ [AU]}\\  
\hline 
1	&	0.83	&	-	&	1.20	&	0.0	&	0.00	&	0.78	&	-	&	1.18	&	0.0	&	0.00\\
5	&	4.81	&	-	&	7.37	&	0.0	&	0.00	&	4.19	&	-	&	6.28	&	0.0	&	0.00\\
10	&	11.90	&	4.28	&	17.40	&	7.6	&	2.13	&	10.30	&	3.03	&	14.4	&	5.7	&	3.13\\
15	&	12.50	&	5.60	&	28.00	&	11.5	&	3.22	&	16.50	&	7.03	&	24.8	&	9.7	&	5.35\\
\hline 
\multicolumn{1}{c|}{ }&\multicolumn{10}{c}{a$_{ini} = 1.5$ [AU]}\\  
\hline 
1	&	0.66	&	-	&	1.04	&	0.0	&	0.00	&	0.60	&	-	&	0.95	&	0.0	&	0.00\\
5	&	4.81	&	-	&	6.02	&	0.0	&	0.00	&	3.42	&	-	&	5.87	&	0.0	&	0.00\\
10	&	9.22	&	1.14	&	13.40	&	3.2	&	0.91	&	7.99	&	-	&	12.50	&	2.8	&	1.55\\
15	&	15.80	&	4.12	&	23.60	&	5.7	&	1.61	&	12.8	&	3.07	&	18.70	&	4.8	&	2.64\\
\hline 
 \multicolumn{1}{c|}{ }&\multicolumn{5}{c||}{}& \multicolumn{5}{c}{}\\
\multicolumn{1}{c|}{ }&\multicolumn{5}{c||}{2.0 M$_{\odot}$, $\mathcal{L}_{0,\star} = 0.61$ [10$^{50}$ g cm$^2$ s$^{-1}$], $t_{rgb} \simeq $ 0.08 [Gyr]}
& \multicolumn{5}{c}{2.5 M$_{\odot}$, $\mathcal{L}_{0,\star} = 0.92$  [10$^{50}$ g cm$^2$ s$^{-1}$], $t_{rgb} \simeq $ 10 [Myr]}\\
 \multicolumn{1}{c|}{ }&\multicolumn{5}{c||}{}& \multicolumn{5}{c}{}\\
\hline 
\multicolumn{1}{c|}{ }&\multicolumn{10}{c}{a$_{ini} = 0.5$ [AU]}\\
\hline 
1	&	0.99	&	-	&	1.30	&	0.0	&	0.00	&	-	&	-	&	-	&	-	&	-	\\
5	&	4.90	&	-	&	7.71	&	0.0	&	0.00	&	3.33	&	-	&	18.4	&	0.7	&	6.82	\\
10	&	11.80	&	8.69	&	16.70	&	14.7	&	19.29	&	7.00	&	-	&	37.9	&	0.7	&	6.82	\\
15	&	16.18	&	13.51	&	22.00	&	19.9	&	26.01	&	7.94	&	-	&	47.0	&	0.7	&	6.82	\\
\hline 
\multicolumn{1}{c|}{ }&\multicolumn{10}{c}{a$_{ini} = 1.0$ [AU]}\\
\hline 
1	&	0.70	&	-	&	0.99	&	0.0	&	0.00	&	-	&	-	&	-	&	-	&	-	\\
5	&	3.60	&	-	&	5.42	&	0.0	&	0.00	&	-	&	-	&	-	&	-	&	-	\\
10	&	7.87	&	-	&	12.10	&	3.8	&	4.98	&	-	&	-	&	-	&	-	&	-	\\
15	&	8.42	&	0.49	&	17.90	&	6.8	&	8.95	&	-	&	-	&	-	&	-	&	-	\\
\hline 
\multicolumn{1}{c|}{ }&\multicolumn{10}{c}{a$_{ini} = 1.5$ [AU]}\\
\hline 
1	&	0.53	&	-	&	0.82	&	0.0	&	0.00	&	-	&	-	&	-	&	-	&	-	\\
5	&	2.82	&	-	&	4.22	&	0.0	&	0.00	&	-	&	-	&	-	&	-	&	-	\\
10	&	6.33	&	-	&	9.18	&	0.9	&	1.14	&	-	&	-	&	-	&	-	&	-	\\
15	&	10.30	&	1.61	&	15.50	&	3.7	&	4.83	&	-	&	-	&	-	&	-	&	-	\\
\hline 
\end{tabular}
\label{table:Omega0.1}	
}
\end{table*}

\begin{table*}
\scriptsize{
\caption{Same as Tab. \ref{table:Omega0.1} but for $\Omega_{ini}$/ $\Omega_{crit}$ = 0.5.} 
\begin{tabular}{c|ccccc||ccccc} 
\hline\hline 
$M_{\rm pl}$ &   v$_{\rm surf}({\rm Be})$ & $\Delta t_{\rm t}({\rm FR})$  & v$_{\rm surf}({\rm Af})$  & $\Delta t_{\rm t+e}({\rm FR})$ & $\Delta t_{\rm t+e}({\rm FR})/t({\rm RG})$ &    v$_{\rm surf}({\rm Be})$ & $\Delta t_{\rm t}({\rm FR})$ & v$_{\rm surf}({\rm Af})$  & $\Delta t_{\rm t+e}({\rm FR})$ & $\Delta t_{\rm t+e}({\rm FR})/t({\rm RG})$\\ 
 $[M_{J}]$ & [km s$^{-1}$]  & [Myr]  & [km s$^{-1}$] & [Myr]  & [$\%$$t_{rgb}$] &  [km s$^{-1}$]& [Myr]   & [km s$^{-1}$] & [Myr]  & [$\%$$t_{rgb}$] \\
\hline 
\multicolumn{1}{c|}{ }&\multicolumn{10}{c}{$\Omega_{ini}$/ $\Omega_{crit}$ = 0.5, Z = 0.02}\\
\cline{2-9}
 \multicolumn{1}{c|}{ }&\multicolumn{5}{c||}{}& \multicolumn{5}{c}{}\\
\multicolumn{1}{c|}{ }&\multicolumn{5}{c||}{1.5 M$_{\odot}$, $\mathcal{L}_{0,\star} = 2.16$  [10$^{50}$ g cm$^2$ s$^{-1}$], $t_{rgb} \simeq $ 0.24 [Gyr]}
& \multicolumn{5}{c}{1.7 M$_{\odot}$, $\mathcal{L}_{0,\star} = 2.39$  [10$^{50}$ g cm$^2$ s$^{-1}$], $t_{rgb} \simeq $ 0.12 [Gyr]}\\
 \multicolumn{1}{c|}{ }&\multicolumn{5}{c||}{}& \multicolumn{5}{c}{}\\
\hline 
\multicolumn{1}{c|}{ }&\multicolumn{10}{c}{a$_{ini} = 0.5$ [AU]}\\
\hline 
1	&	3.21	&	-	&	4.09	&	0.0	&	0.00	&	2.94	&	-	&	3.42	&	0.0	&	0.00	\\
5	&	9.45	&	6.51	&	12.60	&	11.9	&	4.96	&	8.24	&	0.88	&	11.50	&	8.2	&	6.76	\\
10	&	18.60	&	20.92	&	25.50	&	25.1	&	10.51	&	15.60	&	16.20	&	20.80	&	18.3	&	15.09	\\
15	&	28.90	&	 27.52	&	40.40	&	35.9	&	15.00	&	24.00	&	20.78	&	34.50	&	26.8	&	22.15	\\
\hline 
\multicolumn{1}{c|}{ }&\multicolumn{10}{c}{a$_{ini} = 1$ [AU]}\\  
\hline 
1	&	1.69	&	-	&	2.19	&	0.0	&	0.00	&	1.53	&	-	&	2.04	&	0.0	&	0.00	\\
5	&	5.93	&	-	&	8.39	&	0.6	&	0.23	&	5.15	&	-	&	7.23	&	0.0	&	0.00	\\
10	&	14.80	&	7.03	&	20.90	&	9.8	&	4.08	&	11.10	&	3.01	&	15.80	&	5.6	&	4.61	\\
15	&	20.80	&	9.27	&	28.70	&	10.7	&	4.48	&	18.30	&	5.89	&	40.70	&	9.6	&	7.97	\\
\hline 
\multicolumn{1}{c|}{ }&\multicolumn{10}{c}{a$_{ini} = 1.5$ [AU]}\\  
\hline 
1	&	1.15	&	-	&	1.64	&	0.0	&	0.00	&	1.10	&	-	&	1.49	&	0.0	&	0.00	\\
5	&	4.74	&	-	&	6.81	&	0.0	&	0.00	&	3.97	&	-	&	5.84	&	0.0	&	0.00	\\
10	&	9.96	&	1.69	&	15.60	&	3.5	&	1.48	&	8.54	&	0.44	&	12.00	&	2.3	&	1.94	\\
15	&	16.40	&	3.97	&	22.90	&	5.3	&	2.22	&	13.60	&	3.01	&	19.80	&	4.5	&	3.72	\\
\hline 
 \multicolumn{1}{c|}{ }&\multicolumn{5}{c||}{}& \multicolumn{5}{c}{}\\
\multicolumn{1}{c|}{ }&\multicolumn{5}{c||}{2.0 M$_{\odot}$, $\mathcal{L}_{0,\star} = 3.06$  [10$^{50}$ g cm$^2$ s$^{-1}$], $t_{rgb} \simeq $ 0.06 [Gyr]}
& \multicolumn{5}{c}{2.5 M$_{\odot}$, $\mathcal{L}_{0,\star} = 4.65$  [10$^{50}$ g cm$^2$ s$^{-1}$], $t_{rgb} \simeq $ 8 [Myr]}\\
 \multicolumn{1}{c|}{ }&\multicolumn{5}{c||}{}& \multicolumn{5}{c}{}\\
\hline 
\multicolumn{1}{c|}{ }&\multicolumn{10}{c}{a$_{ini} = 0.5$ [AU]}\\
\hline 
1	&	2.87	&	-	&	3.3	&	0.0	&	0.00	&	-	&	-	&	-	&	-	&	-	\\
5	&	5.40	&	-	&	9.6	&	3.6	&	6.19	&	5.20	&	-	&	24.60	&	0.5	&	6.42	\\
10	&	17.10	&	11.62	&	21.3	&	18.6	&	31.86	&	9.22	&	6.42	&	42.1	&	0.5	&	6.42	\\
15	&	18.70	&	16.07	&	33.7	&	26.3	&	45.16	&	12.95	&	6.42	&	57	&	0.5	&	6.42	\\
\hline 
\multicolumn{1}{c|}{ }&\multicolumn{10}{c}{a$_{ini} = 1.0$ [AU]}\\
\hline 
1	&	1.52	&	-	&	1.96	&	0.0	&	0.00	&	-	&	-	&	-	&	-	&	-	\\
5	&	4.72	&	-	&	7.64	&	0.0	&	0.00	&	-	&	-	&	-	&	-	&	-	\\
10	&	8.97	&	0.96	&	12.50	&	3.5	&	6.07	&	-	&	-	&	-	&	-	&	-	\\
15	&	13.80	&	3.81	&	21.30	&	6.1	&	10.39	&	-	&	-	&	-	&	-	&	-	\\

\hline 
\multicolumn{1}{c|}{ }&\multicolumn{10}{c}{a$_{ini} = 1.5$ [AU]}\\
\hline 
1	&	-	&	-	&	-	&	-	&	-	&	-	&	-	&	-	&	-	&	-	\\
5	&	3.39	&	-	&	5.03	&	0.0	&	0.00	&	-	&	-	&	-	&	-	&	-	\\
10	&	7.01	&	-	&	10.20	&	1.3	&	2.22	&	-	&	-	&	-	&	-	&	-	\\
15	&	11.00	&	1.74	&	15.50	&	2.3	&	3.95	&	-	&	-	&	-	&	-	&	-	\\
\hline 
\end{tabular}
\label{table:Omega0.5}
}
\end{table*}

\end{appendix}

\bibliographystyle{aa} 
\bibliography{biblio.bib} 


\end{document}